\documentclass[]{aa} % for the letters 
\def\micron{$\mu$m}
\def\deg{$^\circ$}
\def\arcsec{"}

\def\mdot{$\dot{\rm M}$}
\def\kms{${\rm km}\,{\rm s}^{-1}$\,}
\def\lsun{L/L$_{\odot}$}
\def\rsun{R$_{\odot}$\,}
\def\msun{M$_{\odot}$\,}
\def\msunyr{\msun\,yr$^{-1}$\,}

\usepackage{graphicx}
\usepackage{txfonts}
\usepackage{natbib}

\begin{document}

\title{The yellow hypergiant HR 5171 A: Resolving a massive
  interacting binary in the common envelope phase\thanks{Based on
    observations made with ESO telescopes at the La Silla / Paranal
    Observatory under program ID 088.D-0129 and from Gemini/South
    Observatory under program GS-2011A-C-4.}}

\authorrunning{Chesneau et al.}  \titlerunning{HR 5171 A: a massive
  interacting system in common envelope phase}

 \author{O. Chesneau \inst{1}  \and
         A. Meilland \inst{1}  \and
         E. Chapellier \inst{1}  \and
         F. Millour \inst{1}  \and
         A.M. Van Genderen \inst{2}  \and
         Y. Naz\'e \inst{3}  \and
         N. Smith \inst{4}  \and           
         A. Spang \inst{1}  \and
         J.V. Smoker \inst{5}  \and  
         L. Dessart \inst{6}  \and 
         S. Kanaan  \inst{7}     \and                     
         Ph. Bendjoya \inst{1}  \and
         M.W. Feast \inst{8}  \and
         J.H. Groh \inst{9}  \and    
         A. Lobel \inst{10}  \and        
         N. Nardetto \inst{1}  \and
         S. Otero \inst{11}  \and
         R.D. Oudmaijer \inst{12}  \and
         A.G. Tekola  \inst{8, 13} \and
         P.A. Whitelock \inst{8} \and 
         C. Arcos \inst{7} \and   
         M. Cur\'e \inst{7} \and   
         L. Vanzi \inst{14}} 
          
\offprints{O. Chesneau}

\institute{Laboratoire Lagrange, UMR7293, Univ. Nice Sophia-Antipolis,
  CNRS, Observatoire de la C\^ote d'Azur, 06300 Nice, France\\
  \email{Olivier.Chesneau@oca.eu}%
  \and
  Leiden Observatory, Leiden University Postbus 9513, 2300RA Leiden, The Netherlands 
  \and
  FNRS, D\'epartement AGO, Universit\'e de Li\`ege, All\'ee du 6
  Ao\^ut 17, Bat. B5C, B-4000 Li\`ege, Belgium
  \and
  Steward Observatory, University of Arizona, 933 North Cherry Avenue,
  Tucson, AZ 85721, USA
  \and
  European Southern Observatory, Alonso de Cordova 3107, Casilla
  19001, Vitacura, Santiago 19, Chile 
  \and
  Aix Marseille Universit\'e, CNRS, LAM (Laboratoire d'Astrophysique
  de Marseille) UMR 7326, 13388, Marseille, France
  \and
  Departamento de F\'isica y Astronom\'a, Universidad de Valpara\'iso, Chile
  \and
  South African Astronomical Observatory, PO Box 9, 7935 Observatory,
  South Africa; Astronomy, Cosmology and Gravitation Centre, Astronomy
  Department, University of Cape Town, 7700 Rondebosch, South Africa
  \and
  Geneva Observatory, Geneva University, Chemin des Maillettes 51,
  1290, Sauverny, Switzerland
  \and
  Royal Observatory of Belgium, Ringlaan 3, 1180, Brussels, Belgium
  \and
  American Association of Variable Star Observers, 49 Bay State Road,
  Cambridge, MA 02138, USA
  \and
  School of Physics \& Astronomy, University of Leeds, Woodhouse Lane,
  Leeds, LS2 9JT, UK
  \and
  Las Cumbres Observatory Global Telescope Network, Goleta, CA, 93117,
  USA
  \and
  Department of Electrical Engineering and Center of Astro
  Engineering, Pontificia Universidad Catolica de Chile, Av. Vicu\~na
  Mackenna 4860 Santiago, Chile}

\date{Received ;accepted }

% \abstract{}{}{}{}{} 
% 5 {} token are mandatory
 
  \abstract
  {Only a few stars are caught in the very brief and often crucial
    stages when they quickly traverse the Hertzsprung-Russell
    diagram, and none has yet been spatially resolved in the mass
    transfer phase.}
  {We initiated long-term optical interferometry monitoring of the
    diameters of massive and unstable yellow hypergiants (YHG) with
    the goal of detecting both the long-term evolution of their radius
    and shorter term formation of a possible pseudo-photosphere
    related to proposed large mass-loss events. }
  {We observed \object{HR\,5171 A} with AMBER/VLTI. We also examined
    archival photometric data in the visual and near-IR spanning more
    than 60 years, as well as sparse spectroscopic data. }
  {\object{HR\,5171\,A} exhibits a complex appearance. Our AMBER data reveal a surprisingly large star for a YHG R$_*$=1315$\pm$260\rsun\ (or $\sim$6.1\,AU) at the distance of
    3.6$\pm$0.5kpc. The source is surrounded by an extended nebulosity, and these data also show a large level of asymmetry in the brightness distribution of the system, which we attribute to a newly discovered companion star located in front of the primary star. The
    companion's signature is also detected in the visual photometry,
    which indicates an orbital period of P$_{orb}$=1304$\pm$6d.
    Modeling the light curve with the NIGHTFALL program provides clear
    evidence that the system is a contact or possibly over-contact
    eclipsing binary. A total current system mass of
    $39^{+40}_{-22}$\,\msun and a high mass ratio $q\ge$10 is inferred for
    the system.  }
  {The low-mass companion of \object{HR\,5171\,A} is very close to the
    primary star that is embedded within its dense wind. Tight constraints
    on the inclination and $vsini$ of the primary are lacking, which prevents us from
    determining its influence precisely on the mass-loss
    phenomenon, but the system is probably experiencing a wind Roche-Lobe overflow. Depending on the amount of angular momentum that can
    be transferred to the stellar envelope, \object{HR\,5171\,A} may
    become a fast-rotating B[e]/Luminous Blue Variable (LBV)/Wolf-Rayet star.  In any case,
    \object{HR\,5171\,A} highlights the possible importance of
    binaries for interpreting the unstable YHGs and for massive star
    evolution in general.}

  \keywords{Techniques: high angular resolution; (Stars:) individual:
    \object{HR\,5171A}, \object{V382\,Car}; Stars: binaries: close;
    Stars: circumstellar matter; Stars: massive; Stars: mass-loss }

   \maketitle

%________________________________________________________________

\section{Introduction}
Yellow hypergiants (YHGs) are visually bright evolved sources that
have extreme luminosities of log(\lsun) = 5.6-5.8, which exhibit
episodes of strong mass loss \citep{1998A&ARv...8..145D,
  2009ASPC..412...17O, 2012A&A...546A.105N}. Owing to their large
variability, both photometric and spectroscopic, and 
their spectacular spectral type evolution for some targets, these sources may deserve
to be described as ``Luminous Yellow Variables'' in a similar manner
to the so-called luminous blue variables (LBVs). About ten of
these stars have been clearly identified in our Galaxy so far. In some
cases, uncertainty in their distance estimation renders their
classification as highly luminous sources difficult, and confusion may
exist between classical yellow supergiants (YSGs) and these more
extreme yellow hypergiant (YHGs). Their spectra show many lines that originate in stellar
winds and, in particular, the presence of the infrared Ca~{\sc ii}
triplet, as well as infrared excess from dust
\citep{2013arXiv1305.6051H}. They also show an overabundance of Na
that is supposed to provide indirect evidence of a post-red
supergiant nature, because Na is produced in the Ne-Na cycle in the
high-temperature core of red supergiants. Because of their transitory
nature, YHGs provide a critical challenge for evolutionary modeling,
so they are of great interest for extragalactic studies
\citep{2012ApJ...749..177N, 2012ApJ...750...97D}. 

These stars are also subject to important activity that is characterized by
time variations in the line profiles and striking switches between
emission and absorption in the 2.3$\mu$m CO bands
\citep[e.g., \object{$\rho $\,Cas} on the scale of months as reported
in][]{2006ApJ...651.1130G}. These violent changes are suggested to be
the result of outbursts, which are caused by pulsations occurring
during a period of instability. Occasionally or even permanently, the
wind density is such that it may lead to the formation of a modest-to-strong pseudo-photosphere that might alter the star's apparent
position on the H-R diagram, as seen in Wolf-Rayet stars and
LBVs. \citet{2004ApJ...615..475S} speculate that the YHGs might be
the `missing LBVs' in the Hertzsprung–Russell (HR) diagram, since their true surface
temperature is masked by the screening of the wind.

The instability of YHGs has traditionally been attributed to the star
having the first adiabatic index below 4/3, leading to potentially
strong pulsational activity \citep{2012ApJ...751..151S,
  1995A&A...302..811N}.  \citet{2001ApJ...558..780L} has provided the
theoretical framework of the Ledoux's stability integral $<\Gamma_1>$
atmospheric stability criterion in cool massive stars. Another
possibility, not discussed thus far in the literature, is to invoke
the additional influence of a companion star that may perturb a
loosely bound envelope.

The evolution of such stars across the Herzprung-Russell (HR) diagram
has so far been monitored thanks to photometric and spectroscopic data
taken on a roughly annual basis. The advent of routine optical
interferometry in the southern and northern hemispheres presents an
opportunity to monitor the evolution of the angular diameter of the
YHGs and also their blue counterparts, the luminous blue variables
that experience the so-called S\,Doradus phase
\citep{2009ApJ...698.1698G}. The large variations in the photometric
characteristics and spectral type exhibited by these stars should have
measurable consequences on their apparent diameters. In this
  paper, we focus our work on one source, \object{HR\,5171\,A}. We
  also observed the YSG \object{V382\,Car} (HD\,96918),
  which we used as a reference for the interferometric data on
  \object{HR\,5171\,A}, owing to their similar $K$-band magnitudes,
  1.6 versus 0.9 mag, respectively, and similar expected extension in
  the sky in the range of 2-3\,mas.

\object{HR\,5171 A} (V766 Cen, HD 119796, HIP 67261) is one of the
first objects identified as YHG in our galaxy
\citep{1971ApJ...167L..35H, 1973MNRAS.161..427W, 1992A&A...257..177V}
with a spectral type between G8Ia+ and K3Ia+, but it remains a poorly
studied object. This star and its widely separated B0~Ib companion
\object{HR 5171\,B}, located 9.4\arcsec\ away, dominate the energy
balance in their local environment, causing a large photodissociation
region known as Gum48d \citep{2009ApJ...697..133K,
  2007PhDT........28S}. \object{HR\,5171\,A} lies in the center of the
HII region and was most likely the dominant ionizing
source until its recent post-main-sequence evolution took place. It is
not clear whether the companion is bound at such a large separation, yet
given their apparently similar distances, their isolated location on
the sky, and their short lifetime, it is probable that they both
originate in the same molecular complex. The distance is reasonably
well constrained by several independent estimates to be
D\,=\,3.6$\pm$0.5kpc
\citep{2009ApJ...697..133K,2006AJ....131..603S,1992A&A...257..177V,1971ApJ...167L..35H}. At
such a large distance, the estimated luminosity is log\lsun$\sim$5.7-6,
making \object{HR\,5171\,A} as bright as the famous \object{IRC+10420}
\citep{2009A&A...507..301D,1996MNRAS.280.1062O}.

The cool temperature and the potentially high mass-loss rate of
\object{HR\,5171\,A} have led to the formation of a complex
circumstellar environment. It exhibits a large infrared excess due to
dust with a strong silicate absorption, although no scattered light
was observed around \object{HR\,5171\,A} at visual wavelengths with
deep HST images \citep{2006AJ....131..603S}.
 
 \begin{figure}
 \centering
\includegraphics[width=10.cm]{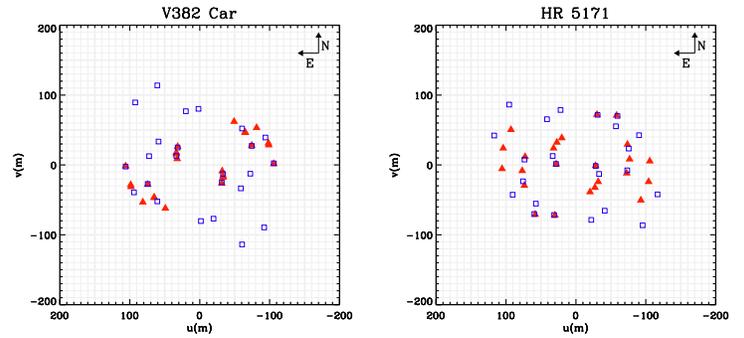}
\caption[]{UV coverage of the AMBER/VLTI observations. Red triangles
  mean medium resolution mode and blue squares low-resolution
  mode.  \label{fig:uv}}
\end{figure}

\begin{figure}
 \centering
\includegraphics[width=8.cm]{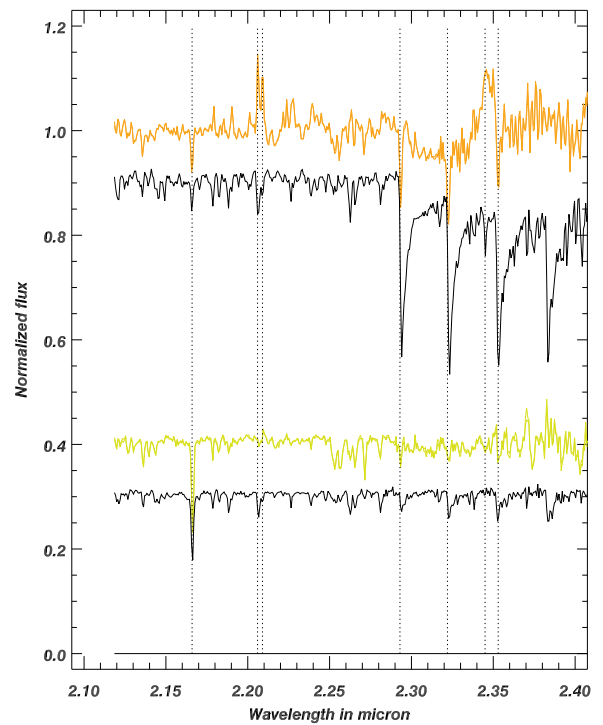}
\caption[]{AMBER/VLTI medium-resolution (R = 1500) K band spectra of
  \object{V382\,Car} (green line, level 0.4) and \object{HR\,5171 A}
  (orange line level 1) compared to a G0Ib and a G8III template
  (levels 0.3 and 0.9, respectively).  Note the striking NaI doublet at
  2.2$\mu$m in the \object{HR\,5171 A}
  spectrum.  \label{fig:specAMBER}}
\end{figure}

Our AMBER/VLTI observations, described below, revealed the extremely
interesting binary nature of \object{HR\,5171 A}, which prompted us to
thoroughly reinvestigate all published photometric and spectroscopic
datasets and to search for possible unpublished data.  The analysis
of such a large dataset cannot be exhaustive so we focus here on
establishing the binary nature of the source by investigating its
spectral evolution and the complex mass-loss history documented by
visual and near-infrared photometry and spectroscopy.

The paper is organized as follows. Section.\ref{sec:obs} describes the
observations, consisting primarily of AMBER/VLTI observations. This
was complemented by archival optical and near-IR data, a few archival
and recent spectra, together with a coronagraphic image of the source
in the near-IR. Sections.3, 4 and 5 focus on the analysis of the
near-IR interferometric, photometric and spectroscopic data,
respectively. Finally, in Sect.\ 6 we discuss the implications of
discovering such a low-mass companion for the fate of this star, and
the potential impact of still undiscovered companions in the yellow or
red supergiant stages.

\section{Observations}
\label{sec:obs}

\subsection{AMBER/VLTI interferometric observations}
\object{V382\,Car} and \object{HR\,5171 A} were observed during 1.5
nights in March 2012 at the Paranal Observatory. The VLTI 1.8m
auxiliary telescopes (ATs) and the AMBER beam recombiner
\citep{2007A&A...464....1P} were used. On March 8 (JD = 2 455 994),
observations were carried out using the low spectral resolution mode
(R=30) and providing a simultaneous record of the $J, H$, and $K$
bands. On March 9 an additional half night of observations was carried
out in medium spectral mode (R=1500) centered at 2.3 $\mu$m and covering
the 2.17 $\mu$m Br$\gamma$ line and the 2.3-2.4 $\mu$m CO bands. All
data were recorded using the FINITO fringe tracker that stabilizes
optical path differences due to atmospheric turbulence. The
fringe tracker allows using a longer exposure time and a
significantly improves the signal-to-noise ratio and overall data
quality. The stars \object{HD\,96566} (spectral type G8III,
$\phi$=1.50$\pm$0.04mas) and \object{HD\,116243} (spectral type G6II,
$\phi$=1.37$\pm$0.04mas) were used as interferometric calibrators and
their diameters obtained from the \texttt{SearchCal} software
\citep{2006A&A...456..789B}. The observations log is presented in
Table\,\ref{AMBERlog} and the ($u,v$) plan coverage for all targets is
plotted in Fig.\,\ref{fig:uv}.  We reduced the data using the standard
AMBER data reduction software {\tt amdlib v3.0.3b1}
\citep{2007A&A...464...29T,2009A&A...502..705C}. The average raw
complex visibility and closure phase were determined using several
selection criteria. The interferometric calibration was then performed
using custom-made scripts described by \citet{2008SPIE.7013E.132M}.

Medium resolution $K$-band spectra were obtained as a byproduct of
the AMBER observations. These spectra are compared in
Fig.\,\ref{fig:specAMBER} with two templates from the
IRTF\footnote{http://irtfweb.ifa.hawaii.edu/$\sim$spex/IRTF\_Spectral\_Library/}
spectral library \citep{2009ApJS..185..289R}.

\subsection{Optical photometry}
\begin{figure*}
 \centering 
\includegraphics[width=17.cm]{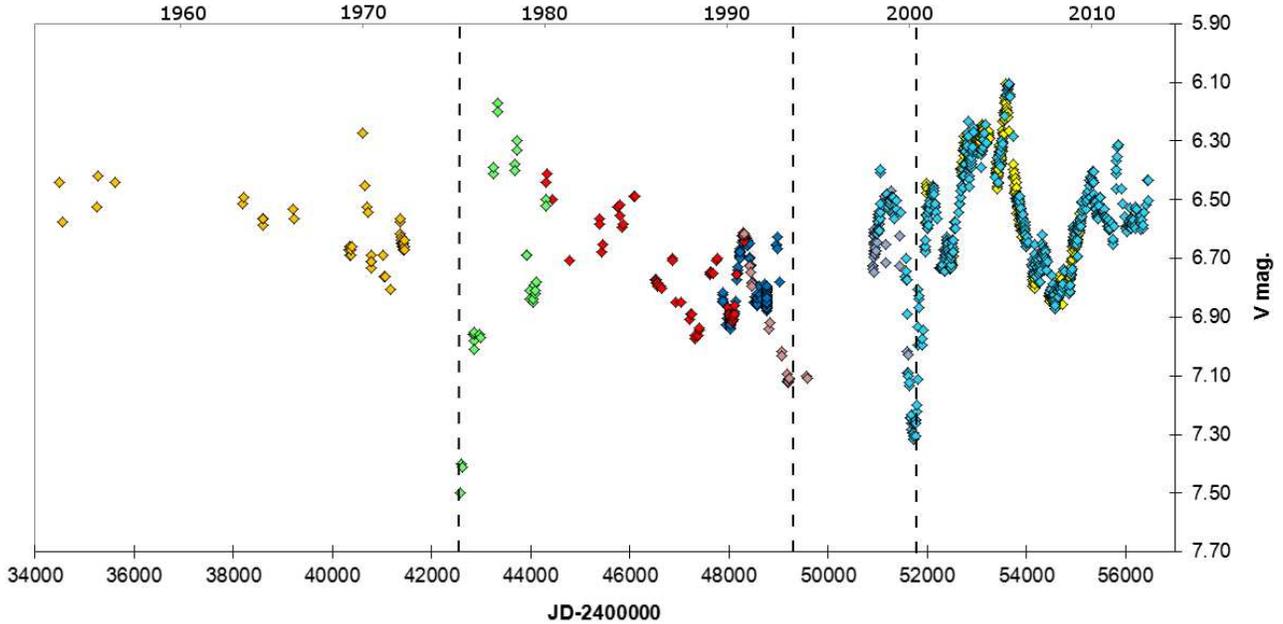}
\caption[]{Visual light curve spanning more than 60\,yr. The colors are
  described in Table\,\ref{tab:photometry}. The low flux events are
  indicated named the Dean, Sterken and Otero minima, which occurred in
  $\sim$1975, $\sim$1994, and $\sim$2000, respectively. \label{fig:lc}}
\end{figure*}

The photometric observations of \object{HR\,5171\,A} are shown in
Fig.\,\ref{fig:lc} and the log of the observations together with some
statistical information can be found in Table\,\ref{tab:photometry}.

The observations consist of several photometric data sets: a
historical one gathered by \citet{1992A&A...257..177V}, consisting of
Johnson $UBV$ and Walraven $VBLUW$ photometry, three unpublished
$VBLUW$ observations made in 1977 (JD 2443248.5, 2443252.5 and 2443269.5, Pel 2013, priv.comm.), Hipparcos (Hp) photometry retrieved
from \citet{1998A&AS..128..117V}, unpublished Stroemgren $uvby$
photomery by the Long-Term Photometry of Variables group (LTPV)
initiated by \citet{1983Msngr..33...10S}, unpublished $V$-band
photometry by Liller, unpublished ASAS-3 photometry
\citep{2002AcA....52..397P}, and recent unpublished $V$-band
photometric observations from Otero. After correcting for the
contribution of \object{HR\,5171\,B}, Liller's $V$-band observations
were made fainter by 0.39 mag to match the $V$ scale. The reason is
probably the use of a non standard filter.

Owing to the extremely red color of \object{HR\,5171\,A} and its high
interstellar and circumstellar reddening, a transformation from the
color index $V-B$ ($VBLUW$) to the $B-V$ of the standard $UBV$ system
is unreliable. It appeared that in order to match the $B-V$ color
indices obtained with a genuine UBV photometer, 0.1 mag should be
added to the computed $B-V$ \citep{1992A&A...257..177V}. No obvious
difference was apparent between $V(UBV)$ computed and $V(UBV)$
obtained with an $UBV$ photometer.  To match the $V(UBV)$
magnitude scale, the Hp magnitudes were corrected by adding 0.15 mag
\citep{1998A&AS..128..117V}. The coincident overlap of part of the
$uvby$ time series with $VBLUW$ and Hipparcos data allowed us to
derive the following corrections: add -0.07 mag to $y$ to get
$V(UBV)$, and add +0.72 mag to $b-y$ to get $B-V$ ($UBV$).

However, all these corrections should be considered with caution.  The
number of overlapping data points obtained in the different
photometric systems is usually small. Additionally, the color
dependence of transformation formulae and of the empirically derived
corrections should
not be underestimated due to the variable color of \object{HR\,5171\,A}.

The temporal sampling of Hipparcos measurements is very
irregular. Therefore we averaged the measurements secured within the
same day, thereby improving the resulting data appearance. Until JD 2453000,
the ASAS-3 light curve suffered from saturation and cannot be used
with confidence, but after this date the scheduling of observations
changed from a single three-minute exposure to three one-minute
exposures. Stars as bright as sixth magnitude became unsaturated. The
agreement with the visual photometry from Sebastian Otero is
excellent, illustrating the quality of these amateur
observations. From the van\,Genderen data, we selected the densest
subset from JD 2446500 to 2448315 during which the mean magnitude of
\object{HR\,5171\,A} was relatively stable. The datasets that included
the $V$ = 10.01 mag companion 9.4\arcsec away were corrected so all
the magnitudes correspond to the G8-K0 hypergiant without the
contribution of the B0 supergiant.

The $V$-band data were complemented by $B$-band
photometry as much as possible. Unfortunately, only sparse $B$-band observations have been
retrieved to complement the extensive dataset published in
\citet{1992A&A...257..177V}. We retrieved a single epoch of
contemporaneous $B$ and $V$-band measurements in the AAVSO archives
(obs: G.\ Di Scala at JD = 2452456.09045).

We also searched for some historical measurements of the visual
magnitude of \object{HR\,5171\,A}. The visual magnitude reported in
1875 by \cite{1879RNAO....1.....G} is 6.8 mag, converted to $V$ = 6.51
mag. This agrees with \citet{1932cdbp.book.....T} who reports
7.0 mag, and with the photovisual magnitude of 6.23 mag from the Henry
Draper catalog \citep{1936AnHar.100....1C}.

\subsection{Infrared photometry}
\begin{figure}
 \centering
\includegraphics[width=9.6cm]{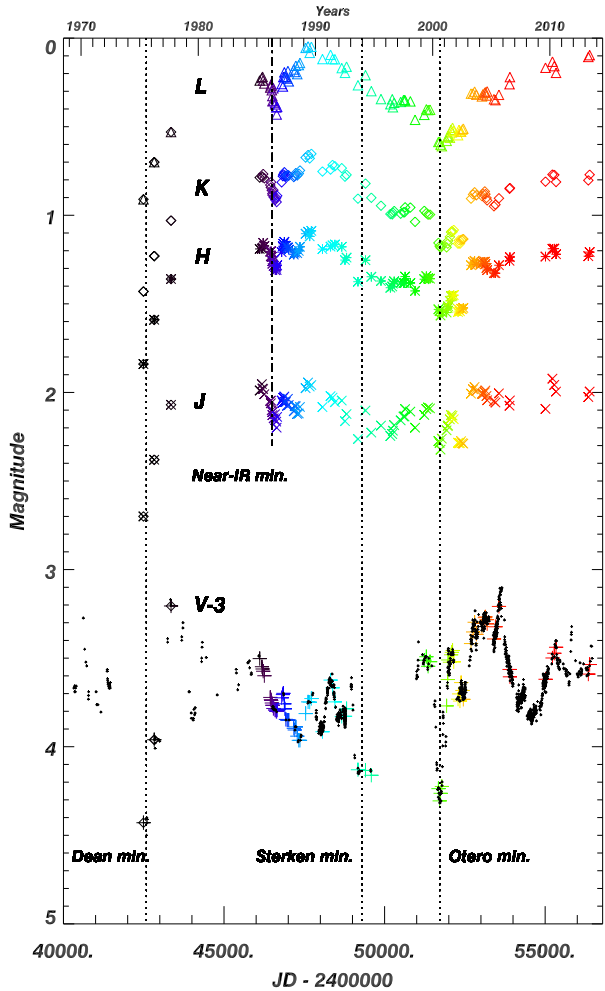}
\caption[]{SAAO near-IR $J, H, K$, and $L$ light curves compared to the
 nearest observations in the V band (the full dataset is shown as small dots). The visual magnitude of \object{HR\,5171\,A} is
    shifted by 3 magnitudes for the sake of
    clarity. Remarkable flux minima are indicated.  \label{fig:lc_nearIr}}
\end{figure}

We present unpublished near-infrared ($JHKL$) data in
Table\,\ref{tab:photometry} and shown in Fig.\,\ref{fig:lc_nearIr}. They were obtained from 1975 until the
present with the MkII infrared photometer, through a 36" aperture, on
the 0.75m telescope at the South African Astronomical Observatory
(SAAO) at Sutherland. The magnitudes are on the SAAO system as defined
by \citet{1990MNRAS.242....1C} and are accurate to $\pm0.03$ mag at
$JHK$ and $\pm0.05$ mag at $L$.  No corrections have been applied for
the contribution of \object{HR\,5171\,B}, which is mostly
negligible. We also present some recent mid-IR data in
Sect.\ref{sec:midIR} in the Appendix.

\subsection{Spectroscopic data}
Visual-wavelength spectra of \object{HR\,5171\,A} are unfortunately
quite rare, although they can help characterize the spectral
type of \object{HR\,5171\,A} and the $v sin(i)$ of the primary star more precisely. High-resolution archival AAT/UCLES spectra taken in 1992 and 1994 were
retrieved from the Anglo-Australian Telescope (AAT)
archives\footnote{{\tiny
    http://apm5.ast.cam.ac.uk/arc-bin/wdb/aat\_database/observation\_log/make}}. They
were taken with UCLES, a cross-dispersed Echelle spectrograph located
at the coud\'e focus of the AAT offering high resolution combined with
good wavelength coverage. The 31.6 lines/mm setting gives almost
continuous wavelength coverage at bluer wavelengths, with a short slit
length (6-15 arcsec).  The data were reduced using IRAF. The quality
of the wavelength calibration of the 1992 spectrum was tested using
nearby observation of the star \object{$\kappa$\,TrA} and was compared
with archival UVES/VLT data. The wavelength calibration of the 1994
data was more difficult to perform because no suitable ThAr reference
could be found in the archive. The 1992 UCLES/AAT spectrum covers the
range 531-810nm, and the two 1994 spectra cover the range
457-697nm. The width of telluric lines has a minimum of 0.0168 nm at
687.8 nm, which is about 7.3\,\kms, providing a lower limit to the
instrumental resolution of R = 41000.

Low- and medium-resolution spectra were acquired at SAAO in 2013 with
the Grating Spectrograph at the 1.9\,m Radcliffe telescope. Several
spectra centered at $\lambda=680nm$ were obtained with gratings 5 and 7 yielding a
spectral resolution of about R\,=\,1000 and R\,=\,6000 over spectral
bands of 370 and 80nm, respectively. The low-resolution spectra were
calibrated with a CuAr lamp and the medium-resolution ones with a CuNe
lamp. A standard star, LTT7379 (G0) was also observed to provide
accurate
spectrophotometry\footnote{http://www.eso.org/sci/observing/tools/standards/spectra/ltt7379.html}.

Radial-velocity monitoring of the source would provide crucial
information about the binary. The only radial velocity data reported
in the literature are from \citet{1982MNRAS.201..105B}. The radial
velocities were obtained by cross-correlating the spectra of several
yellow super- and hypergiants with a G2II template
(\object{$\delta$\,TrA}) and he estimated his error to be 2.5\,\kms. Thirty-eight 
spectra of \object{HR\,5171 A} were secured over 837\,days, between
JD=2443940 (March 7, 1979) and 2444777 (June 21, 1981).
\citet{1982MNRAS.201..105B} reports a strong peak at a period of 263.2
d, and a radial velocity variation of $\sim$9\kms.

We initiated spectral monitoring with Pucheros
\citep{2012MNRAS.424.2770V}, an optical spectrograph of the Pontificia
Universidad Catolica de Chile (PUC). It is a fiber-fed echelle that 
provides spectral resolution R = 20,000 in the visible (390-730 nm)
and a radial velocity accuracy of 0.1\kms. The instrument is installed
at the 50 cm telescope of the PUC Observatory located near Santiago,
Chile (altitude of 1450m). The first observations were executed on
April 11, 2013. Five spectra were taken, each one with 20 min of
integration time, providing a combined spectrum with S/N of about 200.

\subsection{NICI/Gemini South coronagraphic observations}
Observations of \object{HR\,5171\,A} were performed with the NICI
imager \citep{2008SPIE.7015E..49C} on the Gemini South telescope in
February 2011 under program ID GS-2011A-C-4 (P.I. N. Smith). The
dataset consists of a series of 0.4s and 7.6s coronagraphic exposures
with filters probing the narrow $K$-band continuum (Kcont\_G0710,
2.2718$\mu$m, $\Delta \lambda$ =1.55\%) and the [Fe~{\sc ii}] line in
the $H$-band (FeII\_G0712, 1.644$\mu$m, $\Delta \lambda$ =1.55\%). The
Focal Mask F0.9" (G5711) was used. The mask is completely opaque with
a minimum transmission of about one part in 300. Stars centered below
the mask are therefore dimmed by approximately 6.2 magnitudes.

\section{Fitting by geometrical models}
\label{mod:fitting}
\begin{figure*}
 \centering
\includegraphics[width=7.2cm, angle=-90]{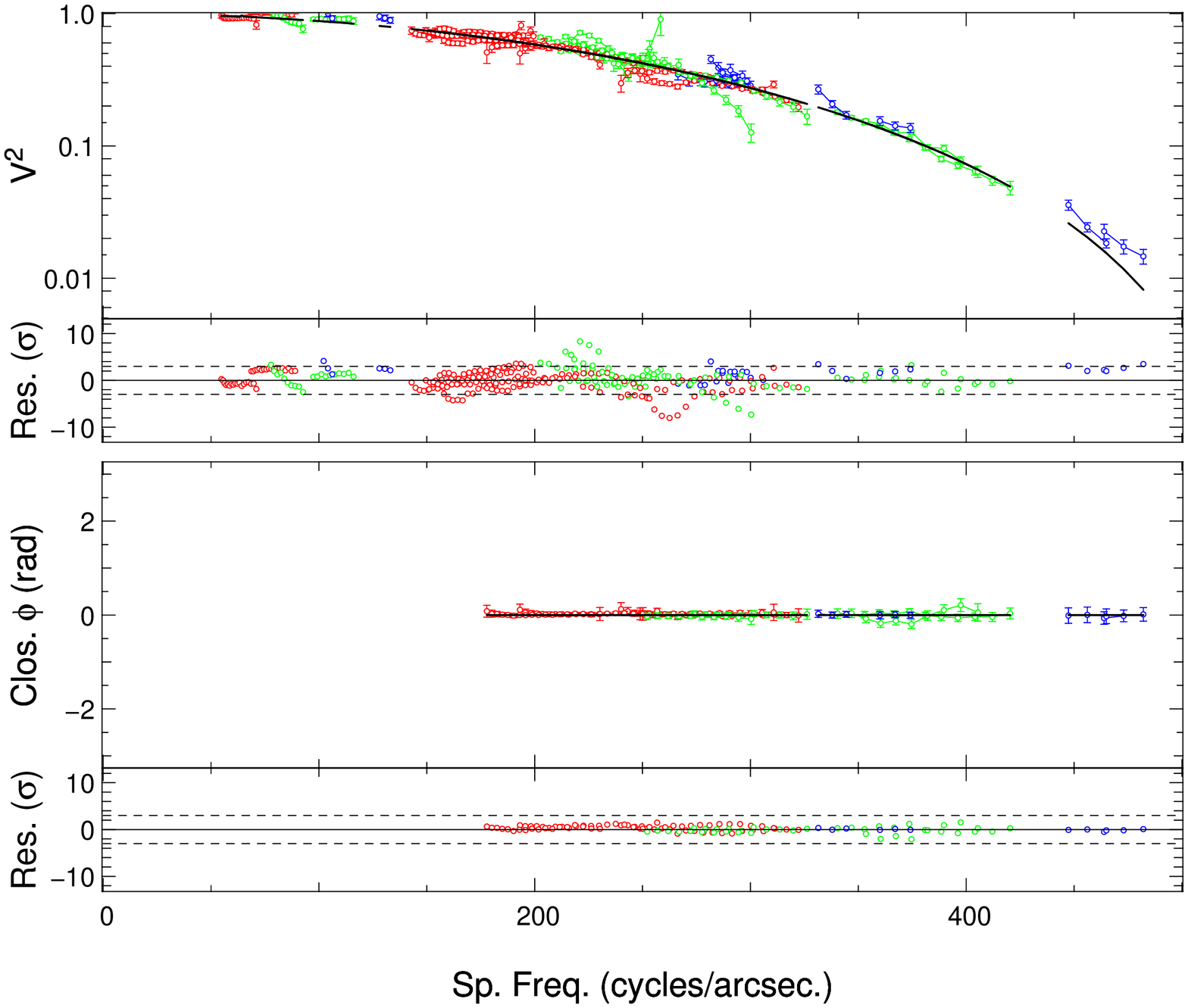}
\includegraphics[width=7.2cm, angle=-90]{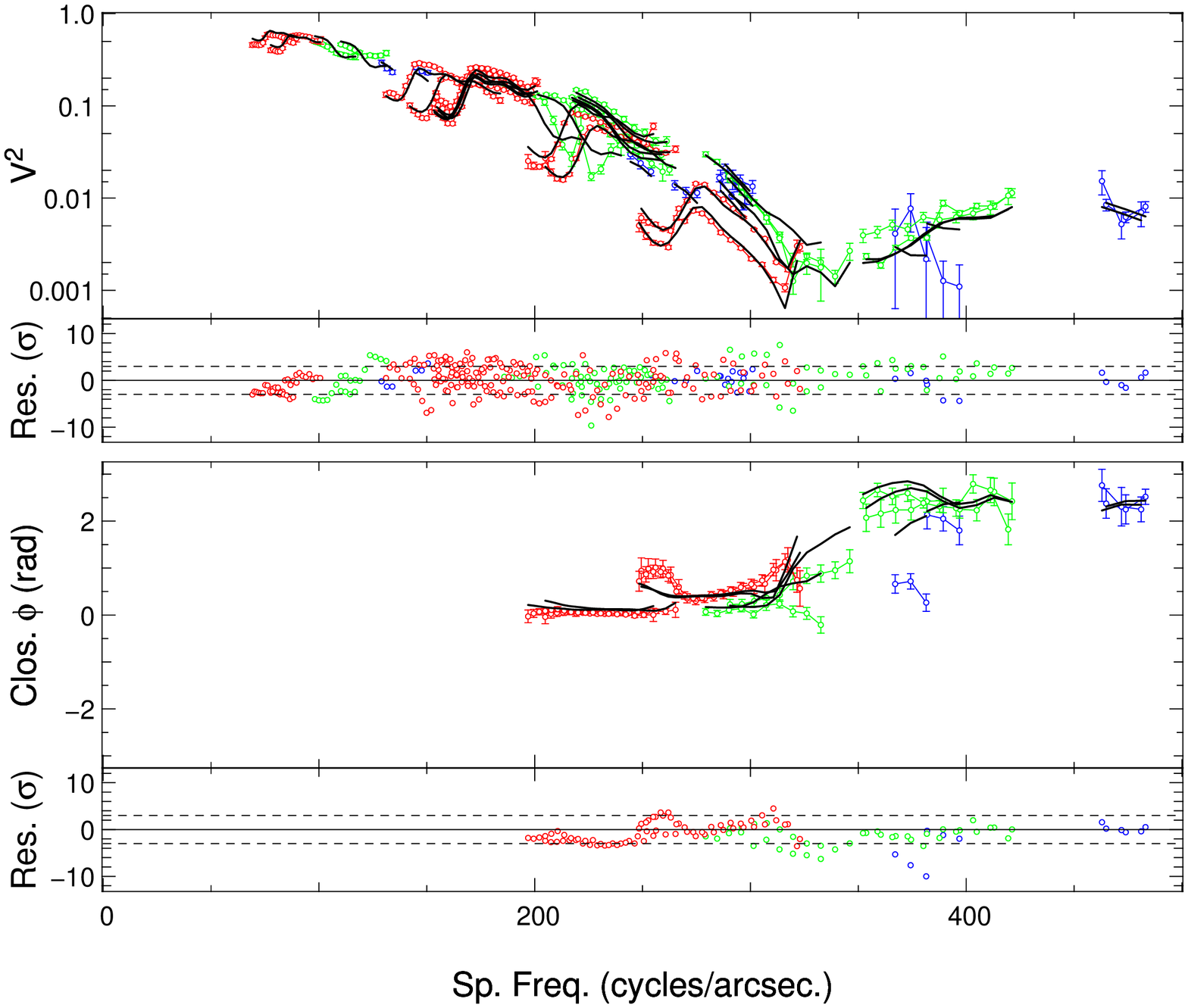}
\caption[]{The AMBER/VLTI datasets of \object{V382\,Car} (left panel) and
  \object{HR\,5171 A} (right panel). For each star, the top panel shows squared
  visibilities in log-scale versus spatial frequencies. Red, green, and
  blue represent data points dispersed in the K, H, and J bands,
  respectively. The residuals are shown below with $\pm$3 sigma
  indicated as dashed horizontal lines. The bottom panel shows the
  closure phases versus the spatial frequencies of the largest of the
  3 baselines involved. The best geometrical models are shown as thick
  and black solid lines. \label{fig:fitomatic}}
\end{figure*}

We analyzed the AMBER data with the {\tt LITpro} software and our own
routine {\tt fitOmatic} \citep[described in the next section and also
in][]{2009A&A...507..317M}. The two packages yield identical results
for \object{V382\,Car}. Both show that the source is best described as
a simple uniform disk (UD) model with a diameter of 2.29\,mas and an
internal accuracy of 5\% (reduced $\chi^2$=3.4). We did not discard
the noisier J band data in the analysis, since it provides important
information at higher angular resolution. We performed some tests on
the presence of diffuse extended emission, or of the existence of
asymmetries in the data. The \object{V382\,Car} data are compatible
with the absence of such emission down to the 6\% flux (accuracy
limit) and an elongation greater than 0.5\% is also excluded. In
Fig.\,\ref{fig:fitomatic} (left), we show an illustration of the data
fit with our best UD model, which can later be compared with
\object{HR\,5171\,A} (right). The interferometric observations
  indicate a lack of significant emission around the star. This can be
  understood if \object{V382\,Car} is a normal YSG 
  rather than a more extreme YHG, since a significant contribution
  from the wind would be expected around a YHG. This classification is
  also consistent with its lack of infrared excess.

The \object{HR\,5171\,A} AMBER dataset shows clear differences
compared to \object{V382\,Car}. 
\begin{itemize}
\item First, the source is larger in angular size, indicated by the
  systematically lower visibilities at the same spatial frequencies.
\item Second, the visibilities at various spatial frequencies deviate 
  strongly from the expected shape of a UD.
\item Third, a closure phase signal is observed rising at high spatial
  frequencies.
\item Fourth, all visibilities and closure phases show significant
  variations with wavelength, which are correlated with spectral
  features, particularly at the end of the K-band owing to the CO first-overtone absorption bands. The source is much more extended at these
  wavelengths due to increased opacity from a thick CO nebula.
\end{itemize}

The medium-resolution data reinforce this view of complex and
spectrally rich information in the dispersed visibilities and
phases. These data are shown and discussed more extensively in the
appendix (Sect.\ref{Sec:MR}). To exploit this spectrally rich dataset, we used our software {\tt fitOmatic}, a
general-purpose model-fitting tool for optical long-baseline
interferometry.  In {\tt fitOmatic}, the gradient descent algorithm is
taken directly from {\tt LITpro}. The main difference with the publicly available tool {\tt LITpro} is the use of wavelength-dependent parameters and a global
optimization method inspired by simulated annealing
algorithms. These two additions allow some flexibility in the fitting
process when dealing with chromatic variations, as in
\object{HR\,5171\,A}. With {\tt fitOmatic}, we test for the
  effect of wavelength-dependent flux ratios between the different
  geometrical models used.

We first tried to fit the \object{HR\,5171\,A} data with simple models
such as a UD or a Gaussian Disk (GD) with a varying radius as a function
of wavelength. Both fits were of poor quality, providing inconsistent
variations in the parameters with wavelength. A fit with a uniform
disk plus a Gaussian Disk, with wavelength-invariant diameters, but
with varying flux ratios provided qualitatively better results for the
visibilities, though the quality of the fit remained poor because none
of these spherical models could by construction account for the
smoothly varying closure phase as a function of spatial frequencies,
which indicates an asymmetric source.

We incorporated this asymmetry in our models by adding an offset
point source with a mean flux fraction that converged to about
  12\% of the total flux at 2.1\micron. Allowing this point source to
be resolved (modeled as a UD) marginally improved the result of the
fit but the angular resolution is insufficient to provide tight
constraints on the secondary diameter apart from an upper limit. Table \ref{fitomatic} lists
all the models tested together with the reduced $\chi^2$ value
obtained.

\begin{figure}
 \centering
\includegraphics[width=8.cm, angle=-90]{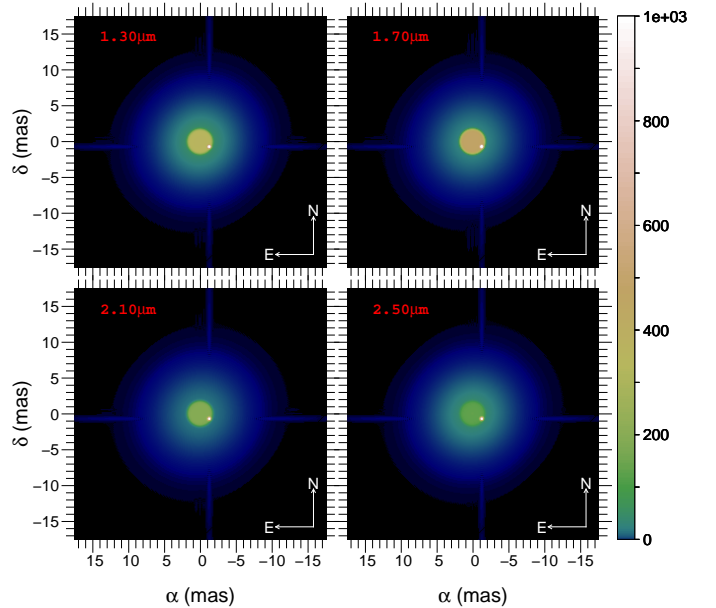}
\caption[]{Interpretation of the near-IR AMBER/VLTI interferometric
  data using a set of three geometrical models. The best model of the
  emitting source consists in a primary photospheric disk
  represented by a uniform disk with a radius of 1.7 mas, a
  circumstellar environment represented by a Gaussian with FWHM of
  4.8 mas, and an unresolved companion represented as a point source
  located at a projected distance of 1.45 mas from the center of the
  primary. The relative fluxes are determined for each spectral channel, representing for instance at 2.1\micron\ 58/30/12\% for the primary,
  circumstellar environment, and secondary,
  respectively.  \label{fig:amber}}
\end{figure}

Therefore, our best interpretation of this dataset (see
Fig.\,\ref{fig:fitomatic}) provides a uniform disk diameter for the
primary of 3.39$\pm$0.2 mas (12.6$\pm$0.5\,AU for a distance
D=3.6\,kpc), surrounded by an extended envelope with a Gaussian full
width at half maximum (FWHM) of 4.76$\pm$0.2 mas (17$\pm$0.7\,AU) in
the continuum. The close companion is visible in the limb of the
primary, separated from the center by 1.45$\pm$0.07 mas (5.15$\pm$0.25
AU), and its flux ratio is 12 $\pm$ 3\% (2.3 mag) of the total flux at 2.1\micron (see the model appearance at
  various wavelengths in Fig.\ref{fig:amber}). The secondary in the model would be smaller
than $\sim$2.0 mas in diameter, but this parameter is weakly constrained with
our restricted $uv$ coverage\footnote{Baselines longer than 150m are
  clearly needed.}.

 Assuming that the source is indeed described well by three components $(i)$ (i.e. the uniform disk, the Gaussian disk and the offset point source with achromatic sizes and positions), 
one can analyze their relative fluxes $C_i$ that were fitted for each independent spectral channel. 
Once these relative fluxes $C_i$ have been obtained, it is straightforward to get the spectrum $S_i$ of each component: 

$$ S_i = \frac{C_i S_\star }{\sum_{j=1}^{n} C_j} $$ 

\noindent where $S_\star$ is the total flux (i.e. the spectrum of the source). We extracted the spectra of the three geometrical components relative to the normalized total flux and discuss the results in Sect.\ref{sec:disc_1}

\begin{table*}
\centering \begin{tabular}{cccccc}
\hline 
Geometrical  & Achromatic & \multicolumn{3}{c}{Flux ratio in percent} & Reduced \\
model & parameters & 1.65\,$\mu$m & 2.10\,$\mu$m & 2.33\,$\mu$m  &$\chi^2$ \\
\hline      
UD+GD& 3.67/4.65 mas  & 80/20 \%  & 50/50 \%  & 35/65\%    & 30 \\
\hline      
UD+GD& 3.39/4.76 mas & 58/35/7  \% & 58/33/9  \% & 30/62/8 \% & 8.6\\ 
+offset PS& $\rho$:1.45 mas, $\theta$=-121\deg\\
\hline
UD+GD& 3.39/4.76 mas & 57/34/9  \% & 58/30/12  \% & 29/60/11 \% & 8.2\\ 
& $\rho$:1.45 mas, $\theta$=-121\deg\\
+offset UD& 1.8\,mas\\
\hline
\end{tabular}
\caption{The different sets of geometrical-models compared to the HR\,5171\,A AMBER/VLTI dataset.  UD, GD and PS stand for uniform disk, Gaussian disk and point source, respectively. $\rho$ and $\theta$ stand for the separation and position angle of the secondary, respectively. The wavelength-dependent parameters are provided for three selected wavelengths and the achromatic parameters (UD diameter, GD FWHM, and PS position) are presented separately. \label{fitomatic}}
\end{table*}

\section{Photometry}
The detection of a companion passing in front of the primary prompted
us to re-investigate the archival visual light curve of this bright
star. \object{HR\,5171\,A} exhibits complex variations in the visible, as shown in Fig.\ref{fig:lc}. The multicolor
photometry was extensively described in several studies and, in
particular, in \citet{1992A&A...257..177V}. A large multicolor dataset
has also been gathered from the blue to the near-IR, see 
Sect.\,\ref{sect:colors}.

\subsection{Light curve analysis}
A statistical analysis of the full visual curve yields a mean at
$V$=6.54 mag and a rms of 0.23 mag. However, this result depends
on the period of observations, and we also performed the same analysis for each dataset. The results are shown in Table\,\ref{tab:photometry}. This simple statistical analysis is important because it reveals several periods lasting several years of either relative quietness or enhanced activities. One can identify several minima, which we
called the Dean ($\sim$2442584, $\sim$1975), Sterken ($\sim$2449330,
$\sim$1993), and Otero ($\sim$2451800, $\sim$2000) minima, separated by $\sim$6716d and $\sim$2470d, respectively. These
minima are all characterized by a magnitude increase to 7.5, 7.2, and
7.3 mag for the Dean, Sterken, and Otero minima, respectively. The
minima lasted about one year but seem to be followed by longer changes, an interpretation supported by the temporal
behavior of the colors (Fig.\,\ref{fig:nearColors}, discussed in
Sect.\ref{sect:colors}). A brightness peak was observed a few years
after the Dean and Otero minima, reaching a magnitude of 6.1-6.2
mag. The recent years probed by the ASAS and Otero observations
(2000-2013) show that \object{HR\,5171\,A} was more active by a factor
$\sim$2 compared to the periods between 1950 and 1970 (Table\,\ref{tab:photometry}), and also
between 1980 and 1992 that appear relatively stable by contrast.

The near-IR light curves shown in Fig.\,\ref{fig:lc_nearIr} provide a
clarification of the $V$-band photometric behavior. The Dean minimum
is very deep and observed in each band, from $V$ to $L$, suggestive of
a cooling of the envelope or/and an important optically thick ejection
of material in the line of sight. The Otero minimum is less marked,
but the imprint of the event can be seen in the $J, H, K$, and $L$ bands
as a sudden 0.2\,mag decrease in magnitude. The Sterken minimum is
only noticeable in the infrared by a small decrease in the flux
($\sim$0.1mag) followed by a rapid return to the the previous
level. We also note that the infrared data suggest at least one minimum in addition to the three described above
\footnote{There might be many much less striking peaks, but is impossible 
to localize them with this limited dataset.} An important decrease of the near-IR fluxes appeared around JD$\sim$2446500 (1986), an event (called 'near-IR minimum' in Fig.\,\ref{fig:lc_nearIr}) associated with a surprisingly weak V minimum.

A long-term evolution on the scale of several thousand days
is observed. The decrease in the $V$-band flux between 1980 and 2000
(see Figs.\,\ref{fig:lc} and \ref{fig:lc_nearIr}) is probably observed $H, K$, and $L$ bands, more significant between 1990 and 2000, but the lack of observations between Sterken and Otero minima prevents a definite answer. An anticorrelation would suggest a varying column density of dust in the line-of-sight causing
absorption in the $V$ band and excess emission in the $L$ band.
The $J$-band photometry exhibits smaller variations with rms of 0.1 mag
compared to rms of 0.15 mag in $K$ and $L$, and larger than 0.2 mag in
the $V$ band.

Turning to the earliest data recorded between 1953 and 1966, we note
that the rms of $V$-band magnitudes reported in
Table\,\ref{tab:photometry} is only 0.1 mag, as low as from
Hipparcos. These two time intervals are relatively very quiet, unlike the current epoch when the source is more active (rms larger than
0.2 mag in the $V$ band).

%\begin{figure}
% \centering
%\includegraphics[width=9.5cm]{flux_HR5171.ps}
% \caption[]{Spectral energy of \object{HR\,5171\,A}. In the optical/near-IR range, two epochs are shown: the average %photometry in dark blue and at the Dean mininimum (1975) in light blue. The photometric points are fitted using two Kurucz spectral with T$_{eff}$=5000K and T$_{eff}$=4200K, for the average and Dean minimum epochs, respectively. The mid-IR domain is simply fitted with a shell at 120\,R$_\star$ with a temperature of 200\,K.   \label{fig:sed}}
%\end{figure}

\subsection{Color analysis}
\label{sect:colors}
\begin{figure}
 \centering
\includegraphics[width=9.5cm]{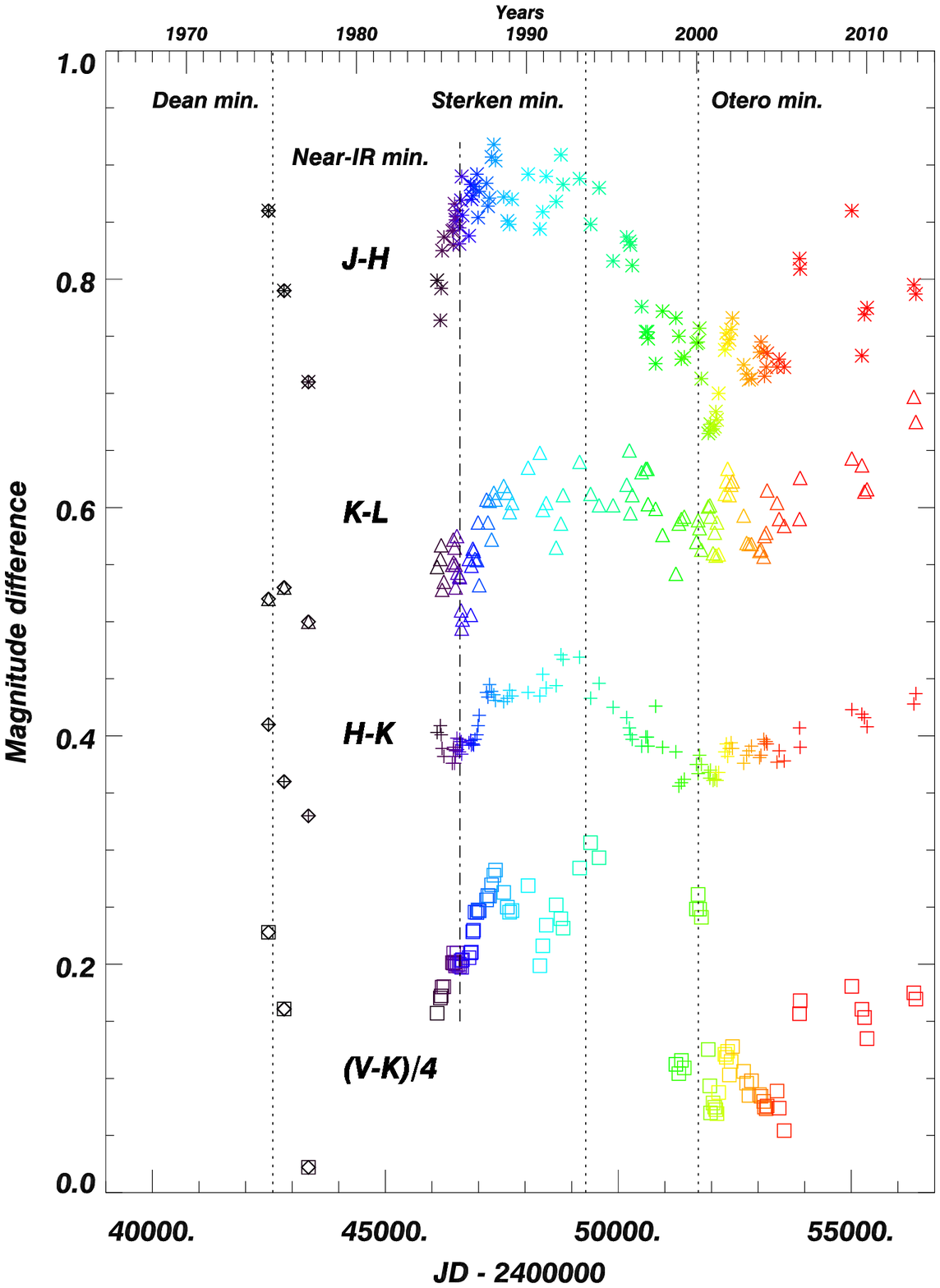}
\includegraphics[width=9.5cm]{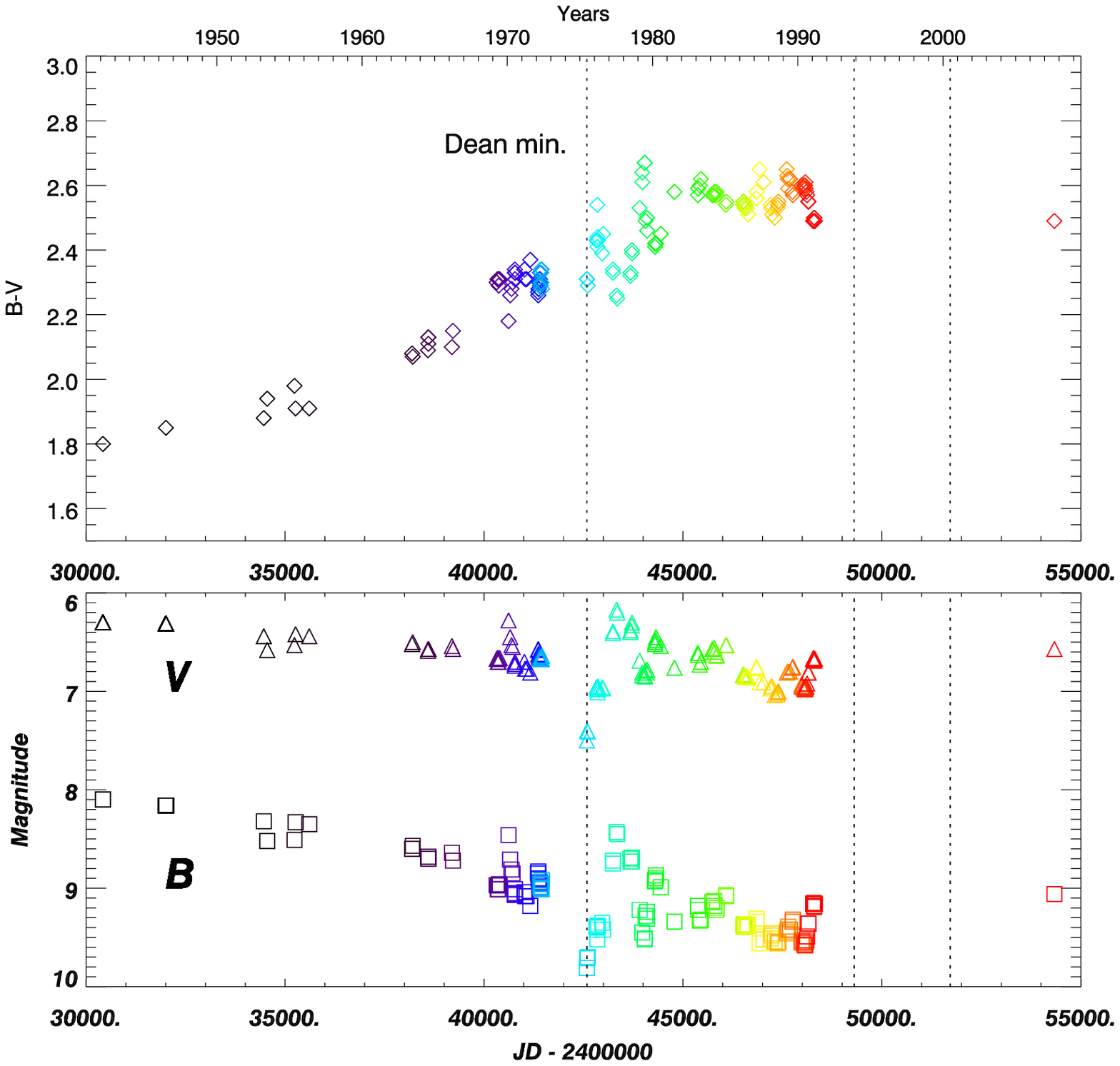}
\caption[]{Temporal evolution of \object{HR\,5171\,A} in several
  colors. The color coding for the curves involving near-IR data is
  the same as in Fig.\,\ref{fig:lc_nearIr}.  The $B-V$ coming 
  from a different dataset are shown separately, together with the corresponding B and V magnitudes. They witness a long-term trend that seems independent of any reddening
  variation. \label{fig:nearColors} \label{fig:BV}}
\end{figure}

The time evolution of near-IR and visible colors indexes is shown in
Fig.\ref{fig:nearColors}. The near-IR colors were computed from the
self-consistent SAAO dataset. The $V-K$ were computed by selecting
several subsets of the visual and $K$-band curves obtained at similar
epochs and interpolating. The $B-V$ dataset is self consistent, mainly
coming from the dataset reported in
\citet{1992A&A...257..177V}. We also constructed two near-IR
  color-color-diagrams, (H-K) versus (J-H) and (K-L) versus (J-K), 
that are shown in Fig.\,\ref{fig:jh-hk} and briefly discussed in
Sect.\ref{sec:midIR}.

The mean color are $<V-K>$=5.78$\pm$0.028 mag, $<J-H>$=0.80$\pm$0.090
mag, $<H-K>$=0.40$\pm$0.027 mag, and $<K-L>$=0.59$\pm$0.039 mag. While
the $V$-band curve appears complex and seemingly erratic, the colors
show a much more coherent behavior.  One can see in
Fig.\ref{fig:nearColors} that $V-K$, $H-K$, $K-L$ and $J-H$ vary with
time in the same manner.  The high rms for the $V-K$ curve probably reflects
the importance of the screening by dusty material in the line of
sight. The rms is minimal for $J-H$ and then increases for $K-L$.

During the Dean minimum, all color indices were very red. Thereafter \object{HR\,5171\,A} became bluer until the high light maximum at JD$\sim$2443300 (the $V-K$,
$H-K$, $K-L$, and $J-H$ are at that time the bluest of all data
points, until now). Then between 1985-1993 the reddening increased,
after which the color indices became bluer again following the
Sterken mimimum and lasting until 2000.  However, a causal
connection between these two phenomena is uncertain.  The Otero
minimum seems to be followed by a period of decreased reddening
lasting up to the present.

In this context, the $B-V$ presented in Fig.\ref{fig:BV} gives an
important insight. The $B-V$ dataset covers more than 60\,yr of data,
and a roughly monotonic temporal evolution is observed
that contrasts with the evolution of the other colors. The curve can
be separated into two main periods: a gradual increase from 1942 to
1982 from $B-V$$\sim$1.8 to $B-V$$\sim$2.6, then an apparent
stabilization, although measurements are unfortunately missing between
1990 and 2010. We also found in \citet{1983TOYal..322...1F} and in
\citet{1958AnCap..20....0J} two $B-V$ indices of 1.8 and 1.85
obtained in 1942.29 and 1946.5, respectively, which prolong the
observed trend (diamonds in Fig.\,\ref{fig:BV}). The curve suggests 
a deep-seated phenomenon that can be interpreted in two ways: a spectral type change or a large variation in
the circumstellar extinction. The curve can be compared 
with the one shown in Fig.\,11 of \citet{2012A&A...546A.105N} which exhibits a similar change in  the B-V for the YHG \object{HR\,8752} but in the opposite direction. This important evolution of $B-V$ is {\it not}
accompanied by an increase in the $V$-band magnitude as would be
expected for increasing circumstellar extinction. Assuming the
classical relationship between $B-V$ and $A_V$ with a factor R=3.1, a
similar trend should be accompanied by a $V$ magnitude increase of at
least 2.2 mag, which is not observed. It also appears that the
1942-1966 light and color curves are more stable than observed
subsequently (see statistics in Table\,\ref{tab:photometry}). The long-term trend of a
$B-V$ increase stops apparently around 1982 and then stabilizes around
$B-V$=2.6 mag for at least 10\,yr, and this is approximately the
current value. Interestingly, the Dean minimum, which appeared as one
of the key events of the past 40\,yr, has a much lower impact on
$B-V$ than the long-term trend. It reinforces our interpretation that
the cause of the gradual $B-V$ evolution is related more to a gradual
change in the underlying star than a variation in the reddening,
even though it is certain that the circumstellar material also has 
strong effects on the color curves that cannot be ignored (see
Sect.\,\ref{sec:env} and Sect.\,\ref{sec:act}). Attributing the full
amplitude of the phenomenon (i.e. $\Delta <B-V>$=0.8 mag) to a
spectral type change implies a significant change of spectral type, from 
G0/G2 to K1/K3. This information is very important in the context of the binary system evolution if one
considers that such a large spectral type change implies an increase (nearly a doubling!) in the radius of the primary
star (Nieuwenhuijzen, Van Genderen et al., in preparation). In this context, one can hypothesize that the relative stability
of the $V$-band magnitudes observed in the 1950s and 60s is a
consequence of the smaller diameter of the primary, hence its lower
sensitivity to the influence of the orbiting companion.

\begin{figure}
 \centering
\includegraphics[width=9.3cm]{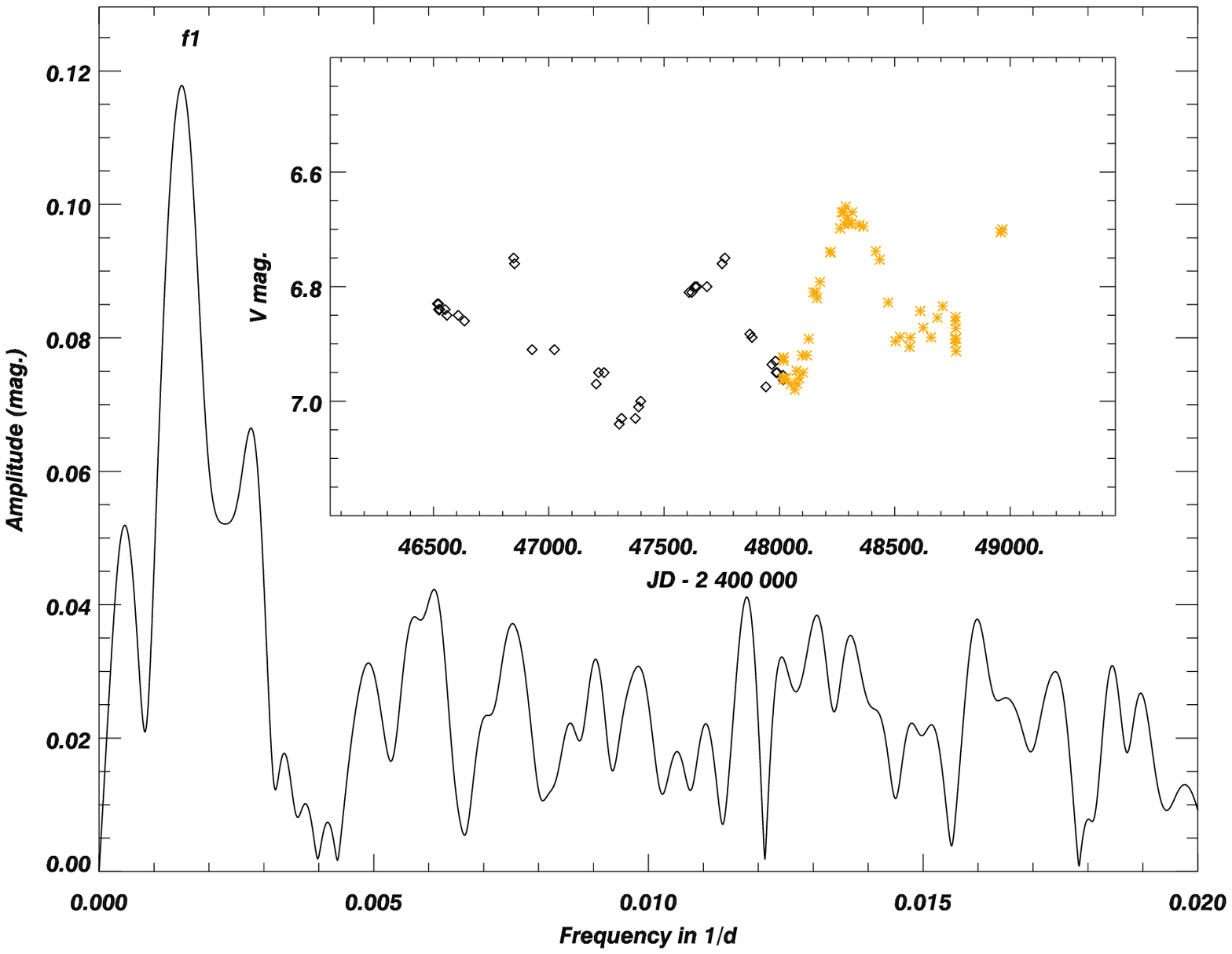}
\includegraphics[width=9.5cm]{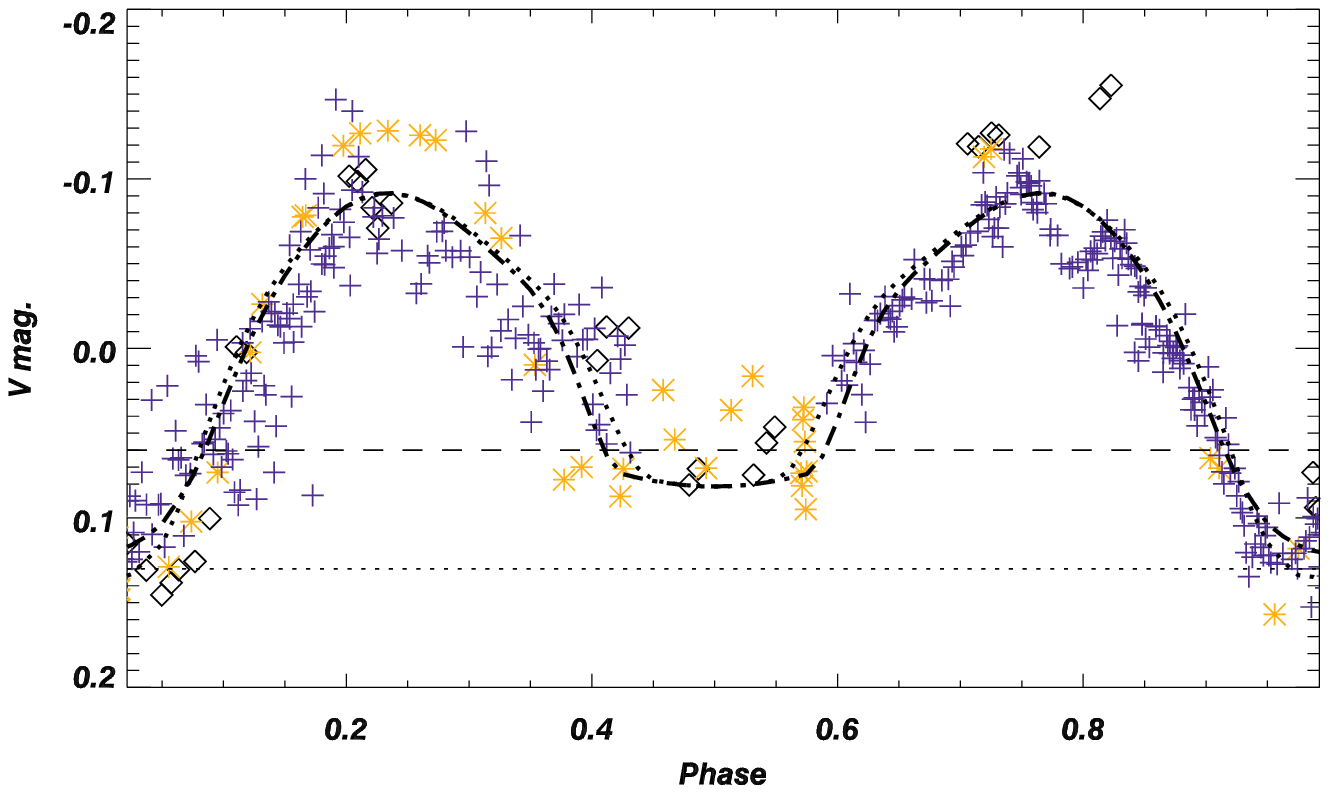}
\caption[]{{\bf Top panel} Density power from the Fourier transform of
  a subset of the visual light curve used for the first search of
  periodical signal. The dataset used, between the Dean and Sterken
  minima, is shown as an inset with the following label: Van Genderen
  data in black diamonds, Hipparcos data in orange stars.  {\bf Bottom
    panel} Phased light curves compared with two {\tt NIGHTFALL}
  models. In addition to the Van Genderen and Hipparcos data (i.e., the most easily prewhitened), the ASAS dataset is shown as blue crosses. The
  dashed horizontal lines are indicative of the deepest level of the
  primary and secondary eclipses, at 0.13 and 0.06 magnitudes, with
  eclipse depth 0.21 and 0.14 magnitudes, respectively. The {\it
    best\,1} and {\it best\,2} models are indicated as dotted and
  dash-dotted curves, respectively. \label{fig:phased}}
\end{figure}

\subsection{Detection of a periodic signal in the $V$ band and the radial velocities}

We performed a Fourier analysis of the visual photometry aiming to
detect a potential periodical signal.  The different sets were
analyzed separately with the {\tt Period04}\footnote{
  http://www.univie.ac.at/tops/Period04/} package
\citep{2005CoAst.146...53L}.

We first used data from the `quiet' period ranging between 1985 and
1992, which have a good temporal coverage and are of good quality. A
clear peak at a frequency of 0.001522d$^{-1}$ ($P_1$=657 d) was detected
first in the Hipparcos dataset (see Fig.\ref{fig:phased}) with an
amplitude a=0.126 mag. For each detected frequency, the amplitude and
the phase were calculated by a least-squares sine fit. We then
enlarged the dataset to include the period of relatively quiet
behavior between the Dean and the Sterken minima (i.e. JD
2443240-2448966). This analysis provides a peak at frequency
0.00152433d$^{-1}$ determined with an accuracy of about 2\%
(corresponding to $P_1$=656$\pm$13d).

Photometry from recent years, including ASAS and Otero data, exhibits more active behavior compared to the Hipparcos dataset. The Fourier
transform of the ASAS data yields a strong peak at
f$\sim$0.00038d$^{-1}$ (P$\sim$3300d, a=0.246 mag) that affects an
accurate determination of the peaks that are present nearby. Three
peaks are also identified at f=0.001542d$^{-1}$ (a=0.074 mag),
f=0.00102d$^{-1}$ (a=0.0393 mag), and f=0.02751 (a=0.0372 mag). We
imposed the f=0.0015d$^{-1}$ frequency in these data set and fitted
the residuals with a little series of sinusoids. Then we came back to
the original observations and prewhitened them with these
sinusoids. The residuals correspond to the cleaned observations. This
prewhitening yields a specific peak around F$\sim$0.0015d$^{-1}$ with
an amplitude of a=0.122 mag, similar to the one found in the 1985-1992
dataset.

It was noticed that if the light curve is phased with the $P_1$
period, an important residual remains at the phases of observed lowest
fluxes. This behavior is particularly clear in the Hipparcos dataset
where the first dip appears lower than the second one. A careful
inspection of the Fourier diagrams revealed a half frequency peak. Phasing the data with the frequency f=0.000761d$^{-1}$ ($P_0$=1314d)
improves considerably the quality of the fit that exhibits a striking
'double-wave' pattern showed in Fig.\ref{fig:phased}. The phased light
curve is reminiscent of an eclipsing system in contact or overcontact
\citep{1994PASP..106..921W}, in line with the finding of the
interferometric observations that the source is seen in front of the
primary. In this case, it is expected that the strong peak detected
(namely $P_1$) corresponds to half the orbital period P$_{orb}$ and
hereafter we will consider that $P_0$=P$_{orb}$.

One may wonder whether the P$_1$ period originates in a slow
pulsation and not from the companion. In many pulsating stars, the
amplitude of the pulsation depends on the wavelength.  If the P$_1$
period were associated to a pulsation we should detect this frequency
dependence in our $B-V$ data. When the $V$ data are fully or
partly analyzed, a frequency peak close to 0.0015 with an amplitude
between 0.1 and 0.12 mag is systematically detected. As a test, such
an analysis was performed on the $B-V$ data. While present
in the $V$ and $B$ datasets, the 0.0015 frequency is missing in the
color index data.

The 38 radial velocities published by \citet{1982MNRAS.201..105B}
cover 1190 days. Structured variations that are significantly larger than the
noise were observed, and the author reports the discovery of a strong
peak at a period of 263.2 days (f=0.003799 d$^{-1}$) from the
dataset. We performed a Fourier analysis with {\tt Period04} that
revealed the presence of two peaks of similar amplitudes at
f=0.003789\,d$^{-1}$ and 0.001467\,d$^{-1}$, the latter one
corresponding to a period P$\sim$682 days, close to P$_1$. Least
squares fits gave K=3.87\kms\ and K=3.96\kms\ for the amplitudes of
the two peaks, respectively, so our analysis suggests that a radial
velocity signal with a period similar to the one inferred from the visual
light curve. In any case, the radial velocities variations imply that the
companion has a low mass compared to the primary or that the
inclination of the system is very low, although the latter hypothesis
is inconsistent with the detection of eclipses.

By comparing the
date of the zero point of the radial velocities with the minimum of
the Hipparcos light curve (JD=2447335$\pm$40), we improved
the precision on the period slightly. We obtained a value P$_1$=652$\pm$3 d,
corresponding to an orbital period P$_{orb}$=1304$\pm$6d. It is worth
noting that the peak corresponding to a period of 263d detected by
\citet{1982MNRAS.201..105B} represents one fifth of P$_{orb}$. In addition, 
\citet{1992A&A...257..177V} identified two periods in their
photometric data: a 430d period during the relatively quiet epochs
ranging from 1969 to 1972 (before the Dean minimum), and then a 494 d
period afterward. These periods were obtained by identifying extrema
in the complex and active light curves, and they are affected by
uncertainties potentially as large as 30d. The harmonic at one third
of P$_{orb}$ has a period of 435d.

The inferred ephemeris with P$_{orb}$=1304$\pm$6d days is given by
$$\Phi  =0.5+(JD-2448000)/P_{orb}$$.

\noindent The light curves covering six orbital periods were used to
test the phasing of the variations. First, a secondary eclipse at
phase $\Phi$ =0.5, when the slightly hotter secondary is in front of
the primary, occurred during the Hipparcos observations at
JD=2448029$\pm$30. Second, the AMBER observations represent a
reference point where the secondary is observed at the beginning or
the end of the secondary eclipse, hence observed near phase
$\Phi\sim$0.4 or $\Phi\sim$0.6 (see Fig.\,\ref{fig:phased}). Using this
ephemeris, the AMBER observations were made at phase 0.63$\pm$0.04,
i.e., at the end of the secondary eclipse. Third, we used the ephemeris
to test and predict the latest maxima of the visual light curve. Some
maxima were identified, but this technique remains highly uncertain
since the light curve is currently severely affected by the activity
of the primary.

\subsection{Modeling of the light curve}
From the phased light curve shown in Fig.\ref{fig:phased}, one can
infer that the distorted surface of the supergiant modulates the
visible flux by 17 $\pm$5\%, and from the mid-period eclipse the
contribution of the companion is estimated to be about 7 $\pm$ 2\% of
the visible flux. This flux contrast is close to the one inferred in the
near-IR by the AMBER/VLTI observation. The emissivity of the companion
is greater than that of the primary, indicating a hotter star. Under
the assumption that the emissive surfaces in the visible and near-IR
are rather similar, this implies that the companion surface
temperature is only hotter by 150-400K than the primary star's
temperature of 3500-4500K. We also note that the secondary eclipse, when
the secondary passes in front, is long, about 0.15P$_0$, implying a
strong inclination for the system.

As a consistency check, we performed an independent analysis based
solely on the phased light curve with the {\tt NIGHTFALL}
code\footnote{http://www.hs.uni-hamburg.de/DE/Ins/Per/Wichmann/Nightfall.html}. This
software \citep{2011ascl.soft06016W} is based on a generalized
Wilson-Devinney method assuming a standard Roche geometry.  {\tt
  NIGHTFALL} is based on a physical model that takes the
non-spherical shape of stars in close binary systems into account, as well as
mutual irradiance of both stars, and it can handle a wide range of
configurations including overcontact systems. We performed a fit of
the light curve by minimizing the free parameters and exploring the
various solutions found. We first assumed a circular orbit and fixed
the period to $P_0$ (i.e., 1333d). We then used various sets of initial
parameters such as a strong inclination (from 45\deg\ up to 90\deg), a
total mass for the system ranging from 30 to 120\,\msun\ and an
identical initial temperature for the two stars of 4000K.  Regardless
of the total mass adopted, the results are very similar. 
%The value of T$_0$ was shifted by 30d compared to the temporal and
%phased original curves. 
The lower mass range is only compatible with large inclinations while
a wider range of inclination accounts for the light curve with a mass
around 120\msun. {\it The full set of solutions favors a primary with
  a large radius (i.e. $\ge$1000\rsun) in contact or over-contact with
  the secondary, involving a filling factor ranging from 0.99 to
  1.04.} Two examples of good models are shown in Fig.\,\ref{fig:phased}
and their parameters are listed in Table\,\ref{Tab:lc}.  The inferred
properties for the primary and secondary are fully consistent with the
limits imposed by AMBER/VLTI measurements. The lower inclinations
close to 45\deg\ yield mass ratios $q$ between 0.8 and 0.5 and a large
temperature difference for the two components (T$_{sec}$/T$_{prim}
\geq 2$). However, such models predict radial velocities for the
primary larger than 20\kms\, which are not compatible with the observed
range of variations, which is less than 10\kms\. Furthermore, a long-duration 
secondary eclipse with a plateau is clearly observed in the phased curve, 
despite the limited S/N, favoring the interpretation with a large inclination.
If one includes
the constraint from the interferometric measurements that the
secondary must have a radius at most one third of the
primary, then a restricted range of parameters is found. A family of
good models involves a high inclination (i$\geq$60\deg), a large mass
ratio ($q\geq$12), and a low-temperature difference between the primary
and the secondary ($\sim$300K).
 
Improved constraints on the system parameters would require more
complete radial velocity monitoring, and we advocate coupled
interferometric and radial velocity monitoring in the future. The {\tt
  NIGHTFALL} modeling of the phased visual light curve is consistent
with the information provided by the interferometric observations,
giving further confidence on the interpretation that the system is a
contact or over-contact binary.

%\begin{figure}
% \centering
%\includegraphics[width=9.3cm]{Best_Radial_Vel_5b_coef1.ps}
%\includegraphics[width=9.3cm]{Best_Radial_Vel_5b_3coef1.ps}
% \caption[]{Radial velocities from \citet{1982MNRAS.201..105B} compared to the light-curve {\tt NIGHTFALL} model {\it best\,1} %($\chi^2=37$, dotted line) and {\it best\,2} ($\chi^2=10$, dashed line)   \label{fig:balona}}
%\end{figure}

\begin{table}
  \caption{Parameter space for the best {\tt NIGHTFALL} models. The period P$_0$=1333d is fixed. \label{Tab:lc}}
\centering \begin{tabular}{ccccc}
\hline 
Param.  & Best\,1 & Best\,2 & min & max  \\
%\hline  
%P & 1333 &\multicolumn{2}{c}{Fixed}  \\    
\hline      
i& 90\,\deg  & 75\,\deg   & 60\,\deg & 90\,\deg\\
Fil.Fac.& 1.03 & 0.99 & 0.98 & 1.04\\
\hline      
Mprim& 60\msun & 74\msun & 20\msun &120\msun  \\
$q$ & 22. & 13. & 7 & 25\\
Msec& 3\msun & 6\msun & 2\msun & 15\msun \\
\hline
Sep$^{\mathrm{2}}$ & 2028 \,\rsun & 2195 \,\rsun  & 1800\,\rsun & 2600\,\rsun\\
L1$^{\mathrm{2}}$ & 1570\,\rsun &  1599\,\rsun & 1400\,\rsun & 1800\,\rsun \\
Prim. Mean radius & 1180\,\rsun  & 1290\,\rsun & 1000\,\rsun&1500\,\rsun  \\
Sec. Mean radius & 312 \,\rsun & 401\,\rsun & 200\,\rsun & 550\,\rsun \\
\hline
Tprim & 4717\,K & 4927\,K & 4400\,K & 5200\,K\\
Tsec/Tprim & 1.02 & 1.06 & 1.01 & 1.25\\
\hline
\end{tabular}
 	\begin{list}{}{}
	\item[$^{\mathrm{1}}$] Separation of the components (circular orbit assumed).
 	\item[$^{\mathrm{2}}$] Position of first Lagrangian point, from primary center
	\end{list}

\end{table}

\section{Spectral analysis}

\begin{figure}
 \centering
\includegraphics[width=9.cm]{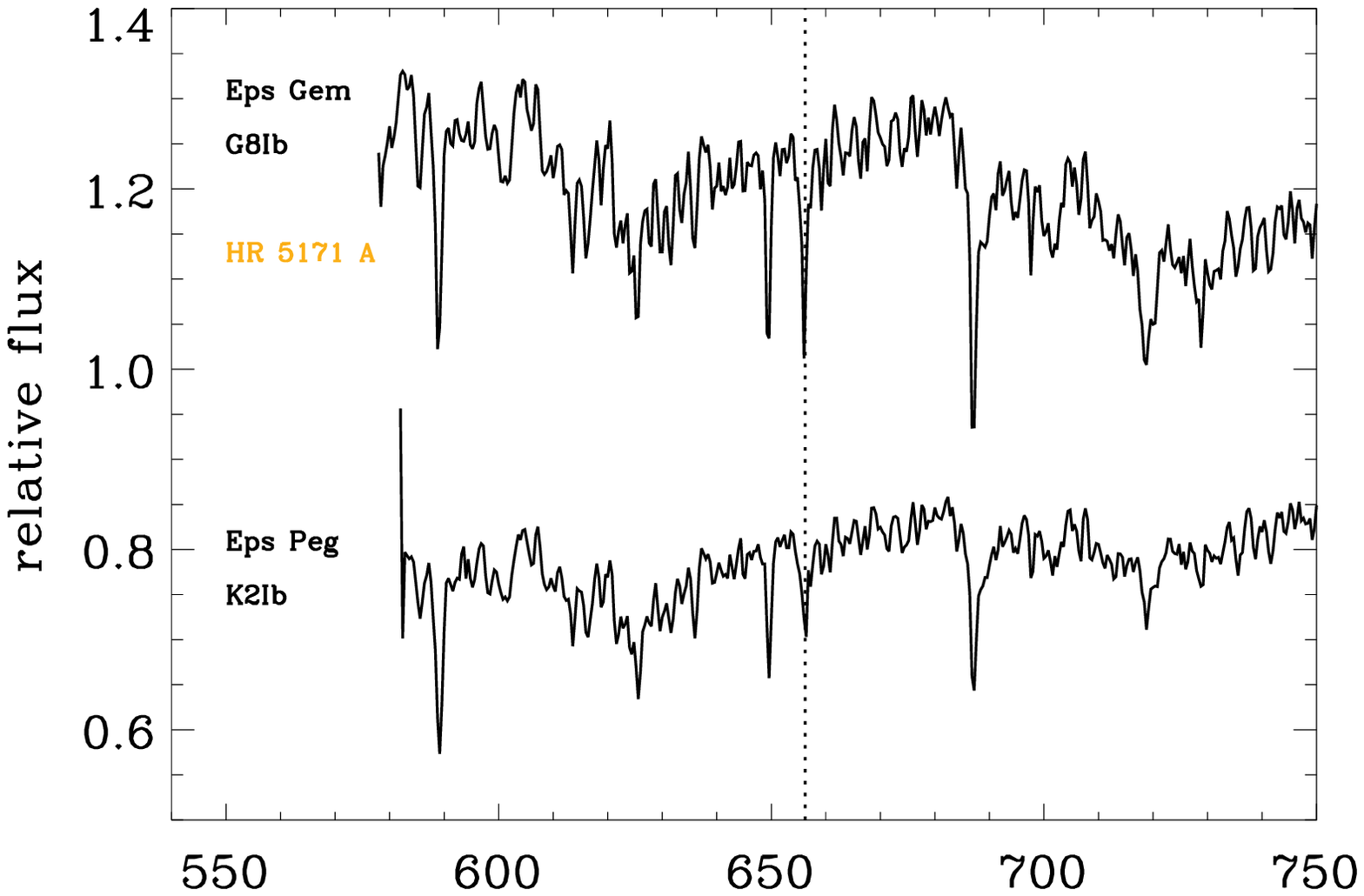}
\includegraphics[width=9.cm]{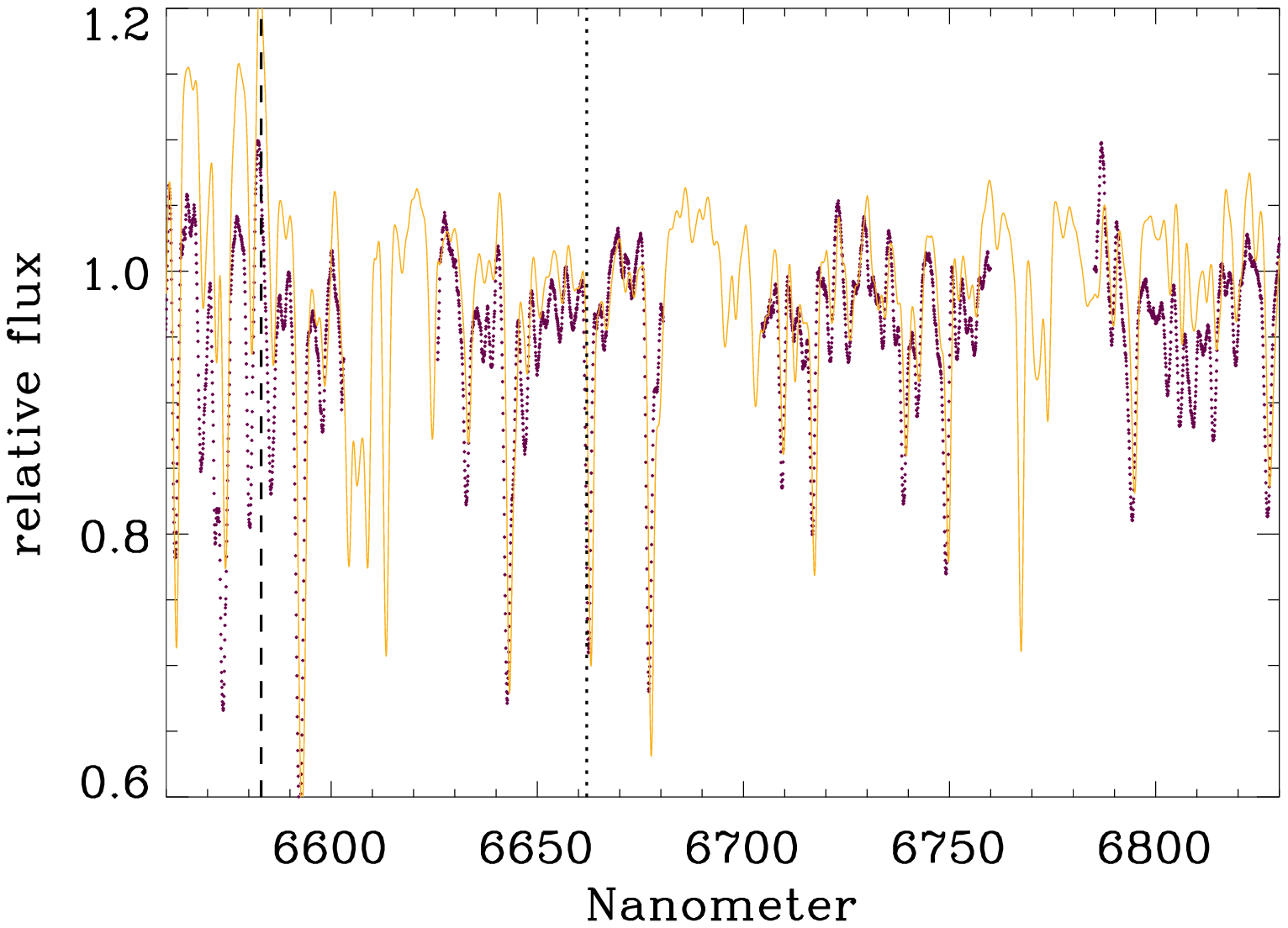}
\caption{{\bf Top}:Comparison of the 2013 low-resolution SAAO spectrum (yellow) with two
  templates of similar spectral type. The spectrum of
  \object{HR\,5171 A} is discrepant in lacking some H$\alpha$
  absorption (vertical dotted line). \textbf{Bottom:} Comparison between the 1992 AAT echelle spectrum (purple dotted line) overplotted with the 2013 PUCHEROS spectrum (yellow curve). The
 spectra are very similar. The H$\alpha$ and  [N\,{\sc ii}] $\lambda$6548 lines are indicated as vertical dotted and dashed lines, respectively.\label{spectra_comp}}
\end{figure}

\subsection{Spectral type}
\label{sec:spec_typ}
The determination of an accurate spectral type using photometric data
for such a variable and reddened object is difficult.  The first
published spectral classification of \object{HR\,5171 A} was reported
by \citet{1971ApJ...167L..35H} to be G8Ia+, but that was corrected by
\citet{1980ApJS...42..541K} to K0Ia+. The spectrum of
\object{HR\,5171\,A} was extensively studied by
\citet{1973MNRAS.161..427W}, providing the first accurate
determination of the effective temperature, T$_{eff}$=4900K. None of
the AAT and 2013 SAAO spectra mentioned above showed any TiO lines. The AAT spectra
secured between 1992 and 1994 are nearly identical. A comparison with
Kurucz models yields a best match with log($g$) = 0 and T$_{eff}$=5000
K for both spectra. The spectra were also compared to templates
providing similar results. We include representative
wavelength regions in Fig.\ref{spectra_comp}.

\subsection{Rotational velocity}

Thanks to the high spectral resolution of his spectra,
\citet{1973MNRAS.161..427W} noticed that the lines are much broader
(50-100\kms) than in an ordinary supergiant ($\sim$8\kms). The
stellar lines in the AAT spectra are obviously resolved well. A simple Gaussian fitting
procedure performed on several lines including the Fe I lines around
600nm (such as 606.5nm) provides FWHM estimates of 49$\pm$9\kms.

However, hypergiant spectra exhibit
broad absorption lines attributed to large scale turbulence motion that may reach
supersonic velocities while the stellar rotational broadening is
considered small \citep{2003ApJ...583..923L, 1998A&ARv...8..145D}. The
discovery of a close-by companion raises a question about on whether the
broadening in \object{HR\,5171 A} has a rotational or turbulent origin.

We therefore performed a Fourier transform (FT) analysis, following
standard techniques \citep{1992oasp.book.....G}. This technique is
considered a robust way to disentangle Doppler broadening from other
sources.  From this analysis, values of $vsini$=57$\pm$15\kms were
estimated from these lines, the large error bar reflecting the rms of
the measurements (the S/N of the AAT spectra is about $\sim$100). A
value of 50\kms, interpreted as the $v sin(i)$ Doppler velocity at
the uniform disk radius inferred from the AMBER observations would 
imply a period of P/sin(i)=1326 days, in fair agreement with our detected period and
derivation of a large inclination. The same technique applied to the
Pucheros spectra at slightly lower resolution but higher S/N 
($\sim$200) provide less scattered results with a mean value of $v
sin(i)$=40$\pm$4\kms. Nonetheless, \citet{2010ApJ...720L.174S} showed that the
determination of the $v sin(i)$ by the FT technique is subject to
important biases when the macroturbulence is significantly greater than
the $vsini$. We used the unblended Fe\,I\,557.2nm line as a good
indicator of variable wind opacity to perform a comparison of the AAT
spectrum of \object{HR\,5171\,A} with a spectrum from
\object{$\rho$\,Cas} and \object{HR\,8752}. \object{HR\,5171\,A}
appears as spectroscopically very similar to these other extreme
stars, raising doubt about our analysis. It is therefore very
difficult to infer this key
quantity from spectroscopy alone. Important is that AMBER/VLTI is able to detect this rotation in
the differential phases without being biased by the pulsation signal
\citep{2004A&A...418..781D, 1995A&AS..109..389C}, provided that the source $v
sin(i)$ is larger than 20 - 30\kms\ (resolving power R = 12000).

\begin{figure*}
 \centering
\includegraphics[width=9.cm]{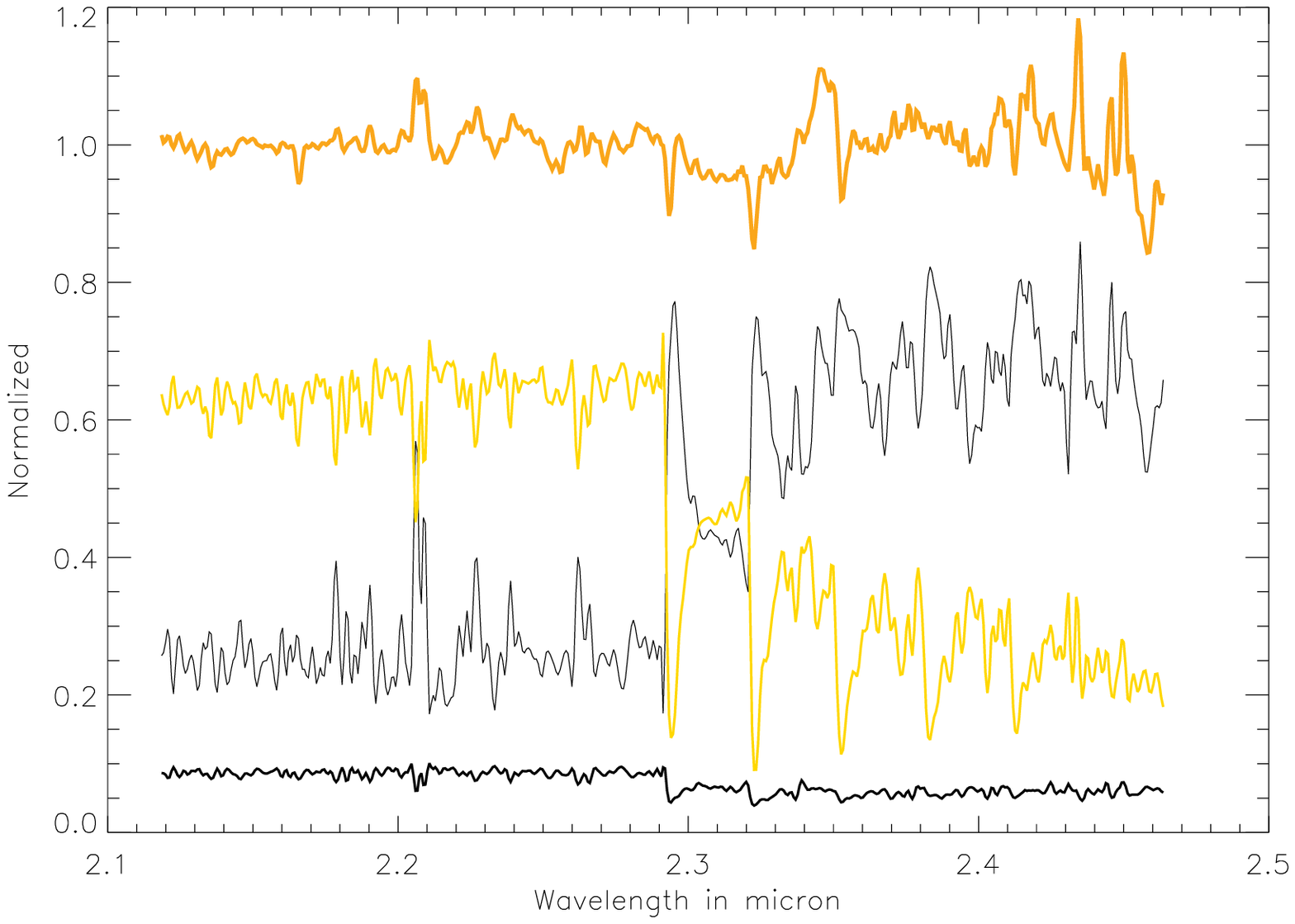}
\includegraphics[width=9.cm]{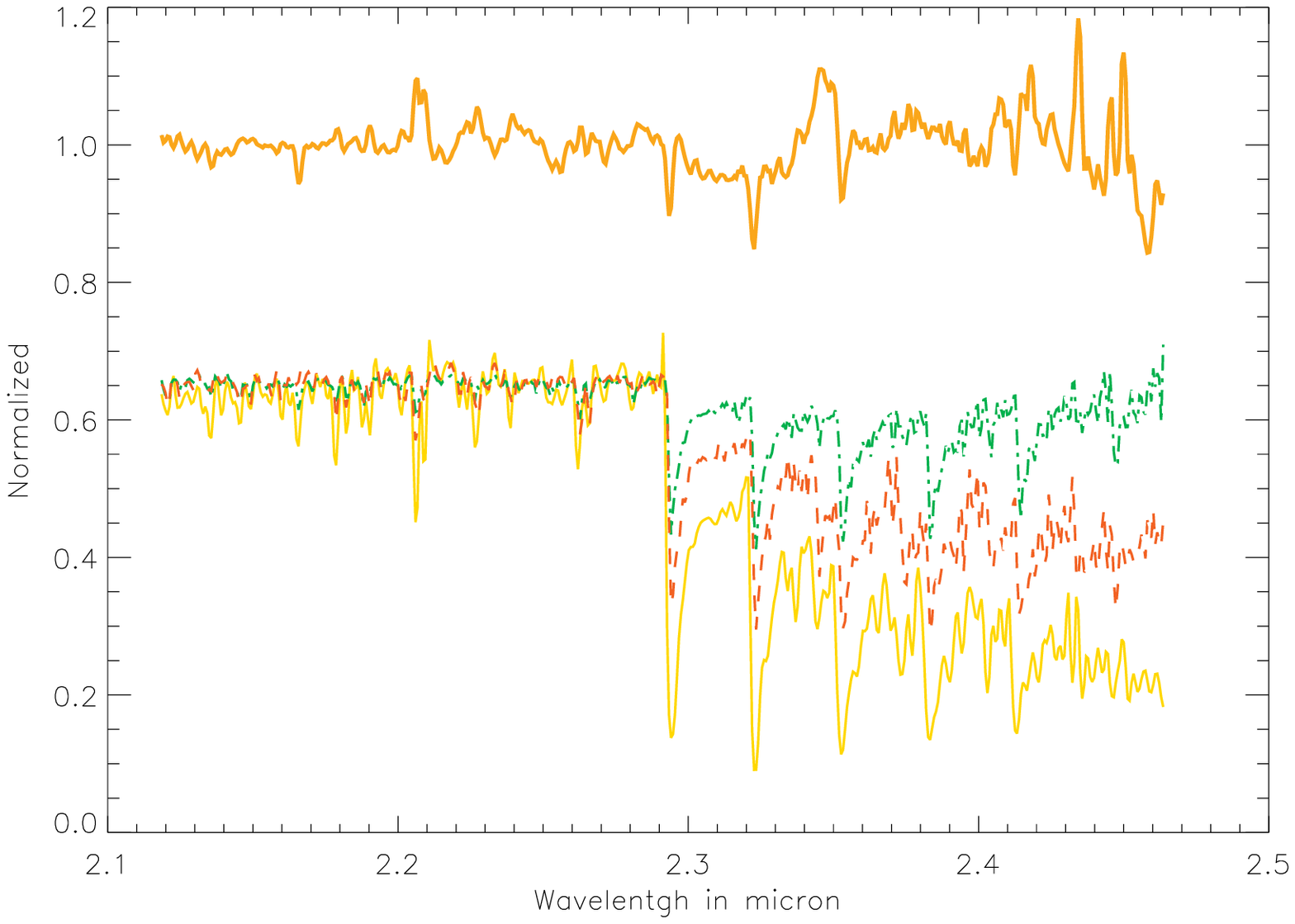}
\caption[]{{\bf Left} Separation of the medium-resolution K band
  spectra \object{HR\,5171 A} based on their spatial properties
  resulting from the geometrical models analysis of
  Sect.\ref{mod:fitting}. The upper orange line is the full
  spectrum. The yellow line is the spectrum from the uniform disk
  which its respective normalized flux. The contribution from the
  Gaussian is characterized by the strong CO emission lines, and the
  secondary flux is the bottom line at about 12\% level in the
  continuum.  {\bf Right} The spectrum from the uniform disk is
  compared to the G8Ib IRTF template in dotted green and the spectrum
  of Betelgeuse in dashed red. The depth of the CO absoptions is
  related to the large column of molecular material that is found
  around the system. Obviously, the CO molecular lines, even
 those originating in the core of \object{HR\,5171 A} are affected by the
  strong veiling from the circumstellar environment and cannot be used
  for the spectral classification of the
  primary.  \label{fig:spec_sep}}
\end{figure*}

\subsection{Influence of the circumstellar environment}
\label{sec:env}
% Section 5.3

In the near-IR spectra shown in Fig.\,\ref{fig:specAMBER}, one can note
the similarity between the G0Ib near-IR template (\object{HD\,185018})
and the AMBER spectrum of \object{V382\,Car}, whereas we observe
numerous differences between the G8Ib template (\object{HD\,208606})
and \object{HR\,5171\,A}. The AMBER spectrum from \object{HR\,5171\,A}
is characterized by a strong sodium line Na\,I\,2.2\,$\mu$m in emission
and narrow CO bandheads that are partly filled by emission.  A similar 
NaI\,2.2\,$\mu$m emission is also reported and discussed by
\citet{2013A&A...551A..69O} on other YHGs, such as
\object{$\rho$\,Cas} \citep{2006ApJ...651.1130G} or
\object{IRC$+$10420} \citep{2013A&A...551A..69O}. This line betrays
the presence of an extended region where the continuum forms inside
the dense wind from the YHG, which veils the hydrostatic photosphere
and may explain the unusual reddening and surprising decoupling
between the photometric and spectroscopic data.

A UD is necessary to explain the zero visibility observed in the
interferometric data that implies the presence of a sharp
(photospheric) border, but the
particular shape of the visibility curve also implies an extended
and dense environment. This envelope is also indicated by the striking
visibility signal observed in the CO lines and the NaI2.2\,$\mu$m
doublet (Fig.\ref{fig:MRspec_Br_vis}), whereas no signal is observed
in the Br$\gamma$ line. A comparison with the same data for
\object{V382\,Car} shows that {\it all} the lines from
\object{HR\,5171\,A} observed in this spectral region form in an
extended environment compared to the continuum, except for Br$\gamma$,
which seems to be the only photospheric line in this spectral
domain.The interferometric model presented in Sect.\ref{mod:fitting} is 
the combination of an uniform disk, a Gaussian disk and an offset point source, 
whose sizes and positions were achromatically estimated in the fitting routine and whose relative fluxes 
are estimated for each spectral channel (see Sect.\,\ref{mod:fitting}). These relative fluxes are shown in
Fig.\,\ref{fig:spec_sep}. The UD spectrum exhibits deep absorptions in
the CO bands, increasing with wavelength more rapidly than in the
spectrum of Betelgeuse, used as template. This suggests that the envelope
renders any spectral classification dubious. The spectrum
from the Gaussian is characterized by the same lines, now
strongly in emission as may be expected from an extended envelope. It must be stressed that although qualitatively interesting, this approach can hardly provide robust quantitative information. The spectral separation procedure relies on an accurate knowledge of the 
flux-calibrated spectrum of the total source at that the time of the observations and the spatial distribution of the emission, implying a full spectrally dependent image reconstruction, such as performed only once on the case of the SgB[e] \object{HD\,62623} \citep{2011A&A...526A.107M}. The formation of CO bands emission is very complex, and certainly their study by the technique described here would imply a (time-consuming)  dedicated imaging campaign. 

We also found that the [N~{\sc ii}] $\lambda$6548 and $\lambda$6583
emission lines of \object{HR\,5171\,A} in the AAT-UCLES spectrum of 14
Jun 1994 are spatially extended. The spatial scale is 0.16 arcsec per
spatial pixel, and the [N~{\sc ii}] $\lambda$6548 line is detected up
to three pixels ($\sim$0.5\arcsec), while the [N~{\sc ii}] $\lambda$6583
line is seen up to five pixels away from the continuum peak
($\sim$0.8\arcsec).  Extended [N~{\sc ii}] $\lambda$6583 emission is
also observed in \object{HD\,168625} and in \object{HR\,8752}
\citep{2013ASPC..470..167L}. It is interesting that both
\object{HR\,8752} and \object{HD\,168625} are supergiant binaries.
Interestingly, HD\,168625 is also an LBV candidate with a triple ring
nebula \citep{2007AJ....133.1034S} resembling the one around SN~1987A.

To summarize, the interferometric data show that the extended envelope
has a strong influence on the spectral appearance, complicating
spectral classification. This envelope acts as a `pseudo-photosphere'
in the sense that it imprints a signature that may have strongly
influenced the spectral diagnoses published so far.

\section{Discussion}

\subsection{\object{HR\,5171\,A}: an over-contact interacting binary}
\label{sec:disc_1}
The first surprise that came out of the AMBER/VLTI observations was
the large angular diameter for such a distant source, which implied a
radius of 1310$\pm$260 R$_\odot$ ($\sim$6 AU). Recent models predict
that the most extended red supergiants reach 1000-1500 \rsun, and have
initial masses not exceeding 20-25 \msun \citep{2012A&A...537A.146E},
while the radius of a YSG is expected to be
400-700\,\rsun. \object{HR\,5171\,A} appears as extended as bright red
supergiants \citep{2013A&A...554A..76A}, such as \object{VY\,CMa}
\citep[1420$\pm$120\rsun]{2012A&A...540L..12W,2001AJ....121.1111S} and
has a radius 50\% larger than the radius of Betelgeuse
\citep[885$\pm$90\rsun]{2009A&A...508..923H}. Such a large radius for
a G8 YHG seems inconsistent with a single star evolution
unless \object{HR\,5171\,A} just left the red supergiant stage.

The phased light curve and the subsequent {\tt NIGHTFALL} modeling
lead independently to a large diameter for the primary and secondary
in the context of a contact or over-contact massive binary system. From
Kepler's third law and for a very-low mass companion of period 3.57\,yr
orbiting as close as 1.3\,R$_*$, we infer the lowest current mass of
the system to be 22$\pm$5 \msun. Taking into account the {\tt
  Nightfall} modeling constraints on the separation of the components and the
optical interferometry constraints on the primary apparent diameter (and their uncertainties), we estimated the total mass of the system to be
$39^{+40}_{-22}$\,\msun (at D=3.6$\pm$0.5kpc).

The long-term evolution of the $B-V$ curve that ended in the
80s suggests that contact may have happened relatively recently,
coinciding with a period of increased activity. Such a large change in
color suggests a dramatic change in spectral type in a few tens of
years, which is not at odds with what is known already for these
extreme objects \citep{2012A&A...546A.105N, 1997MNRAS.292...19K,
  1998A&ARv...8..145D}. Unfortunately, we have to rely on very few
spectra for the spectral type characterization, covering three epochs
1971-1973, 1992-1994, and 2013. The (limited) analysis of the spectra
shows that the spectral type and effective temperature inferred were
relatively similar (a comparison is shown in
Fig.\,\ref{spectra_comp}). The spectral determination remains crude and largely
complicated by the veiling issue. The analysis of the near-IR
spatially resolved spectra from AMBER (Fig.\,\ref{fig:spec_sep}) shows
that the veiling is very strong. An alternative, intermediate scenario
would imply a strong increase in the mass-loss rate, the creation
of a relatively thin, extended gaseous envelope affecting the visual
band only marginally, but strongly influencing the $B-V$ color
index. We discuss this issue further in the following section.

\begin{figure}
 \centering
\includegraphics[width=8.cm]{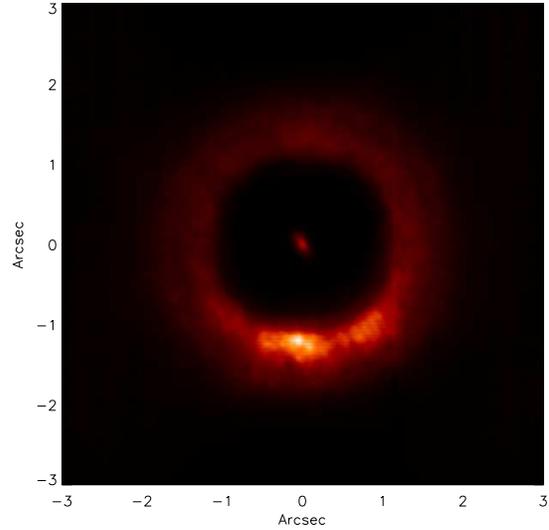}
\caption[]{NICI/Gemini coronagraphic image of HR 5171 A in the [FeII]
  1.644 micron line. The central star is visible, highly dimmed by the
  semi-transparent coronagraphic mask to optimize the fine pointing of
  the source at the center. The mask has a radius of 0.9" and the
  detected light extends up to 1.8".  \label{fig:nici}}
\end{figure}

\subsection{Activity and mass loss}
\label{sec:act}
The photometric behavior of \object{HR\,5171\,A} is very similar to
the activity of the archetypal YHG \object{$\rho$\,Cas}
\citep{2006ApJ...651.1130G, 2003ApJ...583..923L}. The near-IR and
visual light curves show long-term variations and shorter term
minima. If one takes the Dean minimum as the starting point for this
analysis, one can interpret the behavior of the photometric and color
curves as an outburst that propelled part of the primary envelope into
the circumbinary environment. The veiling lasted more than 20 years as
seen by the decrease in the $J-H$ or $V-K$ curves until a new deep
minimum was observed around 2000 (the Otero minimum) that apparently
initiated a new cycle. The Otero minimum is nearly identical to the
millennium outburst $V$-minimum of \object{$\rho$\,Cas}, interpreted
by \citet{2003ApJ...583..923L} as resulting from a shell ejection or
outburst event after which the star continued with its usual pulsation
variability.  The gradual increase in the $B-V$ color index seems to
be only slightly affected by the Dean minimum, suggesting that the
star's activity is a second-order perturbation. Unfortunately, some B
band photometry is missing to probe the later stages (e.g., Otero
minimum).
%\footnote{Keeping in mind that these variations should be observed
%  within a narrower range that the long-term trend.}. 
What is the connection between the short-term deep minima and the
near-IR colors?  During the 1999-2000 outburst, the effective
temperature of \object{$\rho$\,Cas} apparently decreased by more than
3000\,K and an increase of its mass-loss rate from
\mdot$\sim$10$^{-5}$ to $5.4 \times 10^{-2}$\msunyr\ was inferred
\citep{2003ApJ...583..923L}. The analysis of the SED suggests that a
800\,K decrease can explain the differences observed between a minimum
and the normal state.  Intensive spectroscopic and interferometric
monitoring would be needed during the next visual mimimum of
\object{HR\,5171\,A} to better understand the mechanism of these
mass-loss events.

The similarities of the photometric behavior and spectral appearance
of \object{HR\,5171\,A} with other YHG suggest that its activity and
mass-loss rate are due to the same process, namely the chaotic strong
pulsations intimately related to the mean adiabatic index deep in the
atmosphere. {\it The key question related to the discovery of a
  low-mass companion is to determine its influence on the mass-loss
  process. Even a comparatively low-mass secondary can dramatically
  influence a massive --- yet loosely bound and unstable --- envelope
  through tides and atmospheric/wind friction.} As the more massive
star evolves and expands, the Roche lobe limits the size of the
primary, and parts of the envelope become unbound in a process that is
currently poorly known. In particular, it is by no means granted that
this process is smooth and steady. The complex and variable light
curve suggests large ejection events. The primary is forced to have a
surface temperature set by the uncoupled core luminosity and the size
of the Roche lobe.  Binarity may therefore be a key component for
other YHGs as well.

Even with the large dataset presented here, the issue of the rotation
of the primary is far from trivial. We first suspected evidence of
significant rotation from the large widths of lines in the spectrum,
but a comparison to other YHGs suggests that the large turbulence for
these exceptional sources can easily hide a relatively large
(e.g. 10-30\,\kms) $v sin(i)$.  
%A firm conclusion on this side is pending on new high spectral
%resolution AMBER/VLTI observations. 
%A remark in the original Henry Draper catalogue is given as 'narrow
%lines', which may be interpreted as a lower $v sin(i)$ as well as a
%lower turbulence at that epoch \citep[see][]{1936AnHar.100....1C},
%although ``narrow'' is not a quantitative statement. 
We note also that
the presence of the companion is not accompanied by detectable X-ray
emission (see Sect.\ref{Sec:X}). Does this imply that the
mass-transfer process is relatively smooth and the X-rays cannot
emerge from the dense envelope? These observations are correlated to the lack of any
emission in the optical and near-IR from hydrogen recombination lines.

Recent NICI/Gemini-South coronagraphic images obtained in the near-IR
reveal a faint extended nebula with a radius out to 1.8\arcsec
(Rsh=6500\,AU at D=3.6kpc) around the system (Fig.\,\ref{fig:nici}).
This implies that a significant amount of mass has been lost during
past centuries. Using {\it HST}, \citet{2006AJ....131..603S} report 
a lack of diffuse emission in the range 0.9-1.4\arcsec down to very
low levels (5-7.5 mag). Using the assumptions in Fig.\,4 of their paper,
we infer a dynamical time for these radii in the range 300-1000\,yr
(corresponding to expansion velocities of 100 and 35\kms,
respectively). This observation is in line with the estimates from
early studies from \citet{1975ApJ...196..761A} based on the first
mid-IR observations of this source. A model involving a shell at
100\,R$_*$ was considered as not tenable for explaining the IR flux
observed at 25-60\,micron, and a 1000\,R$_*$ shell was preferred by
the authors.  At the distance $D$=3.6kpc, this would represent a shell
of 1.4\arcsec radius, in agreement with the more extended [Fe~{\sc
  ii}] emission seen in the Gemini/NICI images. Does the presence of
the companion imply a long-term cycle with periods of activity and
a high mass-loss rate followed by quiescent periods lasting a few tens
of years, during which time the diameter of the primary is decreased and the
mass-loss rate is much lower?

\begin{figure*}
 \centering
\includegraphics[width=17.cm]{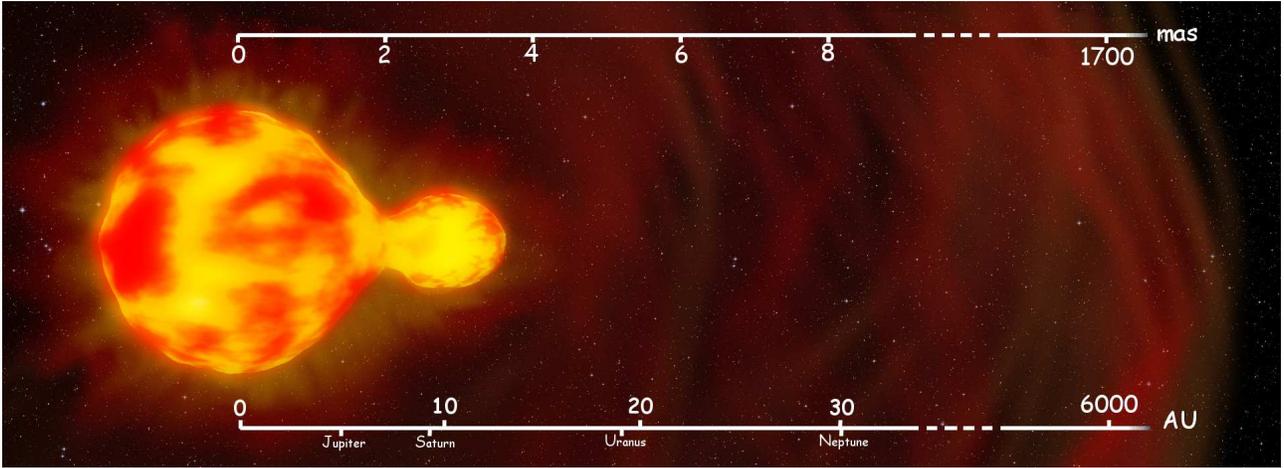}
\caption[]{Roche-lobe model of \object{HR\,5171\,A} seen at phase 0.25
  compared to an angular scale in milliarcsecond (mas) and a distance
  scale in astronomical units (AU) compared with some solar system
  references. The model consists of two stars with a total mass of 39
  solar masses separated by 9.6 AU, an orbital period of 1304d
  (3.57yr). The mass ratio $q$=m$_{sec}$/m$_{prim}$ of 10 was arbitrarily chosen and a
  circular orbit is assumed. The large scales indicate the farthest
  extension of the circumstellar environment. \label{fig:sketch}}
\end{figure*}

In Fig.\,\ref{fig:sketch}, we present a sketch that summarizes our view
of \object{HR\,5171\,A} following our best constraints on the
system. The system is presented at maximum elongation for the sake of
clarity, and observations at these particular orbital phases are
needed to have the best view of the contact between the two components
of the system.

\subsection{Hypergiants and binarity}
Mounting evidence indicates that binarity has a decisive effect on the fates of massive stars and, in particular, on their
rotation rate
\citep{2013ApJ...764..166D,2012Sci...337..444S,2009ApJ...707.1578K}. Binary
mass transfer via Roche lobe overflow (RLOF) has long been considered
a key channel for producing stripped-envelope Wolf-Rayet stars, and
the statistics of supernovae sub-types confirm the importance of the
binary channel in producing stripped-envelope supernovae
\citep{2011MNRAS.412.1522S}.  Similar conclusions are reached through
the modeling of the light curves and spectra of SNe IIb, Ib, and Ic
\citep{2012MNRAS.424.2139D,2011MNRAS.414.2985D,1988ApJ...333..754E}. Despite
the crucial influence of strong binary interaction in the late
evolution of massive stars, very few examples of the phenomenon have
been identified, and this system is, to our knowledge, unique.  Since
the mass transfer phase of massive binaries is very brief
($\sim$10$^4$ yr or so), these systems caught in that phase are
extremely rare and each one of them is very valuable for studying the physics of
the process.

Considering the current parameters for the \object{HR\,5171\,A} system
compared to binary evolution models suggests the following
scenario. During the main sequence, \object{HR\,5171\,A} was a
detached binary system. When the most massive star became a YHG, with a bloated and intrinsically unstable envelope, the
separation was still too large for the two components to merge or to
exchange a large amount of mass. The primary envelope is now quite
cool and large, possibly under the decisive influence of the
companion. The convective layers span most of the star's radius, and
assuming convective motions at 10-20\kms, some material can reach the
interior regions in 2-3\,yr, which is a time shorter than the orbital
period. The Kelvin-Helmholtz time for these stellar envelopes is
comparatively short, a few tens of years. Can the convective envelope
efficiently transfer the incoming angular momentum from the secondary
deep into the primary star's interior? What is the competing influence
of angular momentum loss due to the mass lost by the system? YHGs are
extremely unstable and their observations show that a large
pulsationnal activity can explain their high mass-loss rate even
without invoking the influence of a companion. 

The system is probably undergoing Wind Roche-Lobe Overflow \citep[WRLOF]{2013A&A...552A..26A}, i.e., the primary under-fills its Roche lobe, but a significant fraction of its wind fills it up and is gravitationally channeled to the secondary. This process is by nature non-conservative; i.e., a lot of mass is lost in the process. 
Since the primary is close to Roche-lobe filling, its spin should be tidally synchronized with the orbit rotation (Langer, private communication), but an accurate measurement of
the rotation of the primary is pending. 

The fate of the system depends on the orbital evolution. At present, the orbit could shrink (if some mass flows onto the secondary) or widen (due to the mass lost by the system). If it shrinks, the system could still develop into dynamical common envelope interaction, and the binary could evolve into a short-period WR+x system, where x is either a main sequence stars or a compact star. As such, it would have a small chance of forming a long-duration gamma-ray burst  \citep{2008A&A...484..831D}.  If the orbit widens, it would remain a long period system (Langer, private communication). If the common envelope is avoided but a significant amount of angular momentum transferred, then the primary star could evolve to the blue and become
an LBV spinning close to the critical speed. This provides a possible
way for fast-rotating LBVs such as \object{AG\,Carinae}
\citep{2009ApJ...698.1698G} and \object{HR\,Carinae}
\citep{2009ApJ...705L..25G} to come from binary evolution. \citet{2012ASPC..464..293M} review the binarity of LBV stars, noting
that three of them are in close binaries, including the famous
$\eta$\,Car. A circumbinary disk may also be in formation currently and one can
wonder whether \object{HR\,5171\,A} may evolve to become a
rapidly rotating B[e] star.  The B[e] phenomenon is characterized by
a B-type central star surrounded by a disk at the origin of the large
infrared excess and the forbidden lines observed. This broad definition leads to a heterogeneous group \citep{1998A&A...340..117L}, but despite this, there is a growing amount of evidence that the evolved supergiants stars exhibiting the B[e]
phenomenon (named SgB[e]) are most probably binaries having recently experienced mass-exchange
\citep{2013LNP...857..149M,2013arXiv1305.0459C,2013A&A...549A..28K,
  2012A&A...545L..10W,2012ApJ...746L...2K,2012A&A...543A..77W,
  2009A&A...507..317M}. As an example, \object{HD\,62623} is an
interacting binary harboring an AIab[e] supergiant and a less massive
star ($q\sim$20) orbiting very close in a 150d period. The interaction
is at the origin of the fast rotation of the primary, with a
circumbinary disk of plasma \citep{2011A&A...526A.107M} surrounded at
larger scales by a dense dusty disk \citep{2010A&A...512A..73M}.

Only a few other interacting systems are
known and for those the interaction occurs at much closer separation and the
period is much shorter. Recently, a unique system
harboring two YSGs was detected in a variability survey
of M81. \citet{2008ApJ...673L..59P} discovered it to be an
over-contact binary composed of two YSG orbiting with a
period of P=271d, and they also noticed such a system in the SMC with
F0 supergiants in period P=181d.  The most studied massive system experiencing RLOF is \object{RY\,Scuti} 
\citep{2011MNRAS.418.1959S}, but this interaction occurred at a much
earlier phase of the binary system's evolution since the period is
only 11 days, and the primary is still in a blue supergiant phase. A key difference between these sources is that the envelope of
\object{HR\,5171\,A} is convective. For
\object{RY\,Scuti} the RLOF induces wind focusing and short
mass-transfer events that form a dense disk around the mass gainer
\citep[see their Fig.\,10]{2007ApJ...667..505G}.  \object{HR\,5171 A} is
to our knowledge the first system caught in this stage to have a
resolved common envelope. Interestingly, the strong infrared silicate emission feature in
HR~5171A is most similar to that seen in RY Scuti
\citep{1979ApJ...234L.129G}.  

It is possible that some of the other famous
and unstable yellow hypergiants are experiencing (or have experienced)
a similar binary interaction. \object{IRC+10420} is a yellow
hypergiant assumed to be in a post-red supergiant stage owing to its
large-scale cool envelope, and the star has a current radius estimated
to about 1/3 that of \object{HR\,5171\,A}. The environment around
\object{IRC+10420} is also known to be dusty and bipolar, with
significant equatorial material
\citep{2013A&A...551A..69O,2010AJ....140..339T}. The low inclination
configuration significantly reduces the chances of detecting a faint
periodic signal in the radial velocity or light curves or to detect
rapid rotation. The influence of a putative companion can only be
inferred by careful monitoring via optical interferometry that we
advocate for the few objects of this extreme class. The yellow
hypergiant \object{HR\,8752}, whose appearance is very similar to
those of \object{HR\,5171\,A} is also known to have a hot companion
\citep{2012A&A...546A.105N,1978A&A....70L..53S}. No periodic radial
signature has been reported and the companion's influence may have
been overlooked. The newly discovered Yellow Hypergiant
\object{Hen\,3-1379} (IRAS\,17163-3907) created two large shells,
apparently circular \citep{2013A&A...552L...6H,
  2011A&A...534L..10L}. There is no direct nor indirect evidence of
binarity for this poorly studied source, but this hypothesis deserves
to be carefully checked by high contrast and high angular resolution
observations.

\section{Conclusions}
Using the AMBER/VLTI observations of the YSG 
\object{V382\,Car} as a reference star, we showed that the YHG
\object{HR\,5171\,A} has a very extended photosphere of
1315$\pm$260\rsun\ (at the distance of 3.6$\pm$0.5pc) surrounded by a
diffuse environment. A companion was discovered as a bright spot in
front of the primary disk. Analysis of the visual band light curve
confirmed the eclipsing-binary nature of the system, with a detected
orbital period of of 1304$\pm$6days.

	{\it \object{HR\,5171 A} is an important system caught in the act of
  mass transfer and envelope-stripping. In this particular case, the system
is close enough to us that it can be resolved from ground-based
optical interferometry.}

This system is undoubtedly very interesting and should be investigated
in more detail.  Analysis of the $B-V$ curve suggests that the system
began to interact strongly in the 1980s, following a large increase in
the primary radius. The mass of the system is high with some
uncertainties related to the exceptional configuration of the system
with such a large primary star. Some time will be needed to infer the
orbit from radial velocities coupled to optical interferometric
monitoring. The envelope around the source has a complex shape, and
the interferometric imaging of such a complex target is best performed
with a four-telescope recombiner such as PIONIER/VLTI. We also aim to
better constrain the angular momentum stored in the primary's
extended envelope by better estimating the $v sin(i)$ of the source.
It is extremely difficult to infer this information from spectroscopy
alone, whereas unambiguous information can be obtained using an
interferometer with high spectral resolution capabilities. AMBER/VLTI
is the only instrument of this kind in the southern hemisphere.

Another point that was not studied here is the spatial appearance of
the source in the mid-IR. This wavelength range is of some importance
for determining whether part of the material lost by the system settles
in a circumbinary disk. The MIDI/VLTI instrument is almost decommissioned,
but the second-generation instrument MATISSE/VLTI
\citep{2008SPIE.7013E..70L}, currently in construction phase, will be
perfectly suited to this task.

Some important questions were raised in this study.  Can we observe
the long-term evolution of the diameter of YHGs that are 'bouncing against
the cool border of the Yellow Hypergiant Void' while they are believed
to evolve on a blue loop in the upper H-R diagram? By what factor
does the diameter of a YHG increase during an short-term outburst? Is it related to
the existence of a time-dependent pseudo-photosphere? Interferometry
is bringing a new window for the temporal monitoring of such
stars. Similar monitoring has recently been undertaken in the northern
hemisphere with the VEGA/CHARA visible recombiner
\citep{2009A&A...508.1073M}, complementing the VLTI capability.

\begin{acknowledgements}
  We thank the ESO staff at Garching, Santiago de Chile, and Cerro Paranal
  (Chile), for operating the VLTI. We also thank N. Langer, S. de Mink, O. Absil,
  J.-Ph. Berger, D. Bonneau, J-L Halbwachs, A. Jorissen,
  J-B. LeBouquin, T. Lanz, C. Martayan, F. Martin, H. Nieuwenhuijzen and
  F. Vakili for fruitful discussions. We thank the observers of the
  LTPV group: A. Schoenmakers, J. van Loon, A. Jorissen, O. Stahl,
  S. de Koff, and E. Waelde. J. Manfroid and C. Sterken were
  responsible for the excellent reduction programs. M. Feast and
  P. Whitelock are grateful to the National Research Foundation
  of South Africa for their research grants. We would like to thank
  the following people for contributing infrared observations of
  HR5171: Robin Catchpole, Ian Glass, David Laney, Tom Lloyd Evans,
  Greg Roberts, Bruce Robertson, Jonathan Spencer-Jones, Francois van
  Wyk, and particularly Fred Marang. A. van Genderen is grateful to F. van Leeuwen for many fruitful discussions on the Hipparcos observations
of HR5171A, and to E.H. Olsen for bringing very useful information on the 
$uvby$ photometric system. S. Kanaan and M. Cur\'e acknowledges the support of
  GEMINI-CONICYT project Nº32090006, CONICYT-FONDECYT project Nº
  3120037, CONICYT Capital Humano Avanzado project Nº 7912010046 and
  centro de astrofísica de Valparaíso.  LV acknowledges the support of
  Fondecyt Nº1130849.  SK and LV tanks the PURE team composed from
  N. Espinoza, R. Brahm, and A. Jordan. YN thanks the FNRS and Prodex
  Integral/XMM contracts. This research made use of the Jean-Marie
  Mariotti Center \texttt{SearchCal} and \texttt{LITpro} services
  \footnote{Available at http://www.jmmc.fr} codeveloped by LAGRANGE
  and LAOG/IPAG, and of the CDS Astronomical Databases SIMBAD and VIZIER
  \footnote{Available at http://cdsweb.u-strasbg.fr/}. We acknowledge
  with thanks the variable star observations from the AAVSO
  International Database contributed by observers worldwide and used
  in this research. We thank an anonymous referee for pertinent suggestions that improved the readability of this
    paper.

\end{acknowledgements}

\bibliographystyle{aa}
\bibliography{Bib_YHG} 

%\pagebreak

\clearpage

\appendix

\section{Observation logs}
\begin{table*}
\centering \begin{tabular}{cccccccccc}
\hline \multicolumn{3}{c}{Observation}   & \multicolumn{2}{c}{Projected baseline$^{\mathrm{1}}$}& Mode & DIT$^{\mathrm{2}}$ & CT$^{\mathrm{3}}$ &Seeing & Object \\
\hline      Date& Time          & Triplet  & B(m)          & P.A. ($^o$)                 &      & (s) & (ms)             & (")   & \\
\hline

2012-03-08&03:03:47&K0-A1-G1&128.7/73.9/81.7&-132.1/84.0/-164.4 &LR-K-F$^{\mathrm{4}}$&0.05&4.5&1.37&V382 Car\\
2012-03-08&04:10:17&K0-A1-G1&129.0/67.7/80.3&-152.0/60.4/-178.8 &LR-K-F&0.05&6.6&0.90& HR 5171\\
2012-03-08&04:52:51&K0-A1-G1&124.2/79.1/77.4&-109.8/107.2/-147.8&LR-K-F&0.05&9.6&1.00&V382 Car\\
2012-03-08&05:31:29&K0-A1-G1&128.4/73.4/79.4&-134.1/80.2/-165.6 &LR-K-F&0.05&9.5&1.01& HR 5171\\
2012-03-08&06:58:20&I1-A1-G1& 99.9/79.7/35.6&-64.8/134.1/-111.1 &LR-K-F&0.05&16.5&0.85&V382 Car\\
2012-03-08&07:46:53&I1-A1-G1&106.0/79.2/40.2&-88.8/110.1/-128.3 &LR-K-F&0.05&17.2&1.02& HR 5171\\
2012-03-08&08:34:07&I1-A1-G1& 92.0/78.1/28.4&-40.3/156.6/-93.0 &LR-K-F&0.05&17.0&0.80&V382 Car\\
2012-03-08&09:20:49&I1-A1-G1&101.9/80.0/35.8&-67.3/130.7/-111.2 &LR-K-F&0.05&22.1&0.99& HR 5171\\
2012-03-09&05:05:13&I1-A1-G1&105.9/79.5/41.1&-89.1/110.6/-129.8 &MR-K-F&1.00&7.8&0.72&V382 Car\\
2012-03-09&05:28:39&I1-A1-G1&105.7/73.6/43.6&-118.5/80.8/-152.4 &MR-K-F&1.00&6.4&0.81& HR 5171\\
2012-03-09&06:11:02&I1-A1-G1&102.7/80.0/37.9&-73.9/125.3/-118.0 &MR-K-F&1.00&7.1&0.67&V382 Car\\
2012-03-09&06:38:46&I1-A1-G1&106.7/77.2/42.2&-103.0/96.2/-139.9 &MR-K-F&1.00&6.9&0.69& HR 5171\\
2012-03-09&07:27:36&I1-A1-G1& 97.3/79.2/33.3&-56.7/141.6/-105.2 &MR-K-F&1.00&5.1&0.88&V382 Car\\
2012-03-09&07:50:38&I1-A1-G1&105.8/79.4/39.9&-87.0/111.8/-126.9 &MR-K-F&1.00&6.7&1.04& HR 5171\\
2012-03-09&08:55:19&I1-A1-G1&103.1/80.0/37.0&-72.4/125.9/-115.1 &MR-K-F&1.00&5.0&1.20&V382 Car\\
2012-03-09&08:34:23&I1-A1-G1& 91.8/78.1/28.1&-39.4/157.3/-92.3 &MR-K-F&1.00&6.7&1.05& HR 5171\\
\hline
\hline
\end{tabular}
\caption{Observations log of V382\,Car and HR\,5171\,A. \label{AMBERlog}}
 	\begin{list}{}{}
	\item[$^{\mathrm{1}}$] Projected baseline length B and position angle P.A.
	\item[$^{\mathrm{2}}$] Detector integration time
	\item[$^{\mathrm{3}}$] Atmospheric coherence time
 	\item[$^{\mathrm{4}}$] LR-K-F: AMBER low-resolution mode (R = 30) in the K band, using the fringe tracker. MR-K-F: same with medium-resolution mode (R = 1500)
	\end{list}

\end{table*}

\begin{table*}
\centering 
\caption{Description of the visual light curve datasets. The color and symbol table are used to generate Fig.\,\ref{fig:lc}. \label{tab:photometry}}
\begin{tabular}{cccccccccccc}
\hline 
\hline 
Reference  & Data & \multicolumn{2}{c}{First data}   &  \multicolumn{2}{c}{Lastest data}  &  \multicolumn{6}{c}{V magnitude} \\
\hline 
  & Nb. & MJD$^{\mathrm{1}}$ & date & MJD$^{\mathrm{1}}$ &date & Coding  & median & mean & rms & min & max \\
\hline 
Harvey$^{\mathrm{2}}$ & 54 & 34497 & {\tiny 1953-04-28} & 41460 & {\tiny 1972-05-22} & {\tiny Orange}  & 6.64 & 6.62&0.10 & 6.27 & 6.80 \\
\hline
Dean$^{\mathrm{3}}$ & 30 & 42584 & {\tiny 1975-06-20} & 44299 & {\tiny 1980-02-29} & {\tiny Green} & 6.81 & 6.75 & 0.34 & 6.17 & 7.50\\
\hline 
Van Genderen$^{\mathrm{34}}$ & 64 & 43248 &  {\tiny 1977-04-14}& 48315 &  {\tiny 1991-02-27} & {\tiny Red} & 6.75 & 6.72 & 0.15 & 6.41 & 6.97\\
\hline
Hipparcos$^{\mathrm{5}}$ & 326 & 47871 &  {\tiny 1989-12-10}& 49011&  {\tiny 1993-01-23} & {\tiny Dark blue} & 6.84 & 6.81 & 0.09 & 6.59 & 6.93\\
\hline
LTPV group$^{\mathrm{6}}$ & 21 & 48283&  {\tiny 1991-01-26} & 49568&  {\tiny 1994-08-03} & {\tiny Brown}&
7.08 & 7.00 &  0.19 & 6.67 & 7.17\\
\hline
Liller$^{\mathrm{7}}$ & 45 & 50899&  {\tiny 1998-03-26} & 51602&  {\tiny 2000-02-27} & {\tiny Gray} & 
6.22 &    6.22 &  0.13  &    6.05 &    6.65\\
\hline
ASAS$^{\mathrm{8}}$ & 351 & 52980&  {\tiny 2003-12-06}  & 55088 &  {\tiny 2009-09-13} & {\tiny Yellow} & 6.61 & 6.56 & 0.21 & 6.11 & 6.87\\
\hline
Otero$^{\mathrm{9}}$ & 510 & 50989&  {\tiny 1998-06-24}  & 56331&  {\tiny 2013-02-07}  & {\tiny Light blue} & 6.59 & 6.61 & 0.22 & 6.10 & 7.32\\
\hline
\hline 
& & & &  &  &  \multicolumn{6}{c}{J magnitude} \\
\hline 
SAAO &109&42491& {\tiny 1975-03-20}&56390& {\tiny 2013-04-07} & &2.11  &   2.09 & 0.11 &   1.92 &   2.70 \\
\hline 
& & & &  &  &  \multicolumn{6}{c}{H magnitude} \\
\hline 
SAAO & 109& // & //&// &// && 1.31   &   1.27 &   0.14   &  1.09 &    1.84\\
\hline 
& & & &  &  &  \multicolumn{6}{c}{K magnitude} \\
\hline 
SAAO & 109& // & //&// &// & &0.91 &     0.89 &     0.15 &   0.65 &    1.43\\
\hline 
& & & &  &  &  \multicolumn{6}{c}{L magnitude} \\
\hline 
SAAO & 106& //&// &// & // & & 0.33 &     0.31 &    0.16 &    0.05 &    0.91\\
\hline 
\end{tabular}
 	\begin{list}{}{}
	\item[$^{\mathrm{1}}$] MJD = JD - 2400000
 	\item[$^{\mathrm{2}}$] From \citet{1972MNSSA..31...81H}, included in \citet{1992A&A...257..177V}
	\item[$^{\mathrm{3}}$] From \citet{1980IBVS.1796....1D}, included in \citet{1992A&A...257..177V} 
	\item[$^{\mathrm{4}}$] From \citet{1992A&A...257..177V}, including three unpublished VBLUW observations made in 1977 (JD 2443248.5; 2443252.5 and 2443269.5, Pel, 2013, priv.comm.)
 	\item[$^{\mathrm{5}}$] Hipparcos data published in \citet[see Sect.2.2]{1998A&AS..128..117V}.
	\item[$^{\mathrm{6}}$] Private communication to van Genderen from `Long-Term Photometry of Variables at La-Silla' group \citep{1983Msngr..33...10S,1991A&AS...87..481M,1993A&AS..102...79S} 
	\item[$^{\mathrm{7}}$] Private communication to van Genderen from W. Liller. Observations performed with a CCD and a 
20 cm private telescope (see Sect.2.2). 
	\item[$^{\mathrm{8}}$] From http://www.astrouw.edu.pl/asas/ \citet{2002AcA....52..397P}
	\item[$^{\mathrm{9}}$] This study, http://varsao.com.ar/Curva\_V766\_Cen.htm
	\end{list}

\end{table*}

\section{Medium-resolution AMBER/VLTI data}
\label{Sec:MR}
\begin{figure*}
 \centering
\includegraphics[width=17.cm]{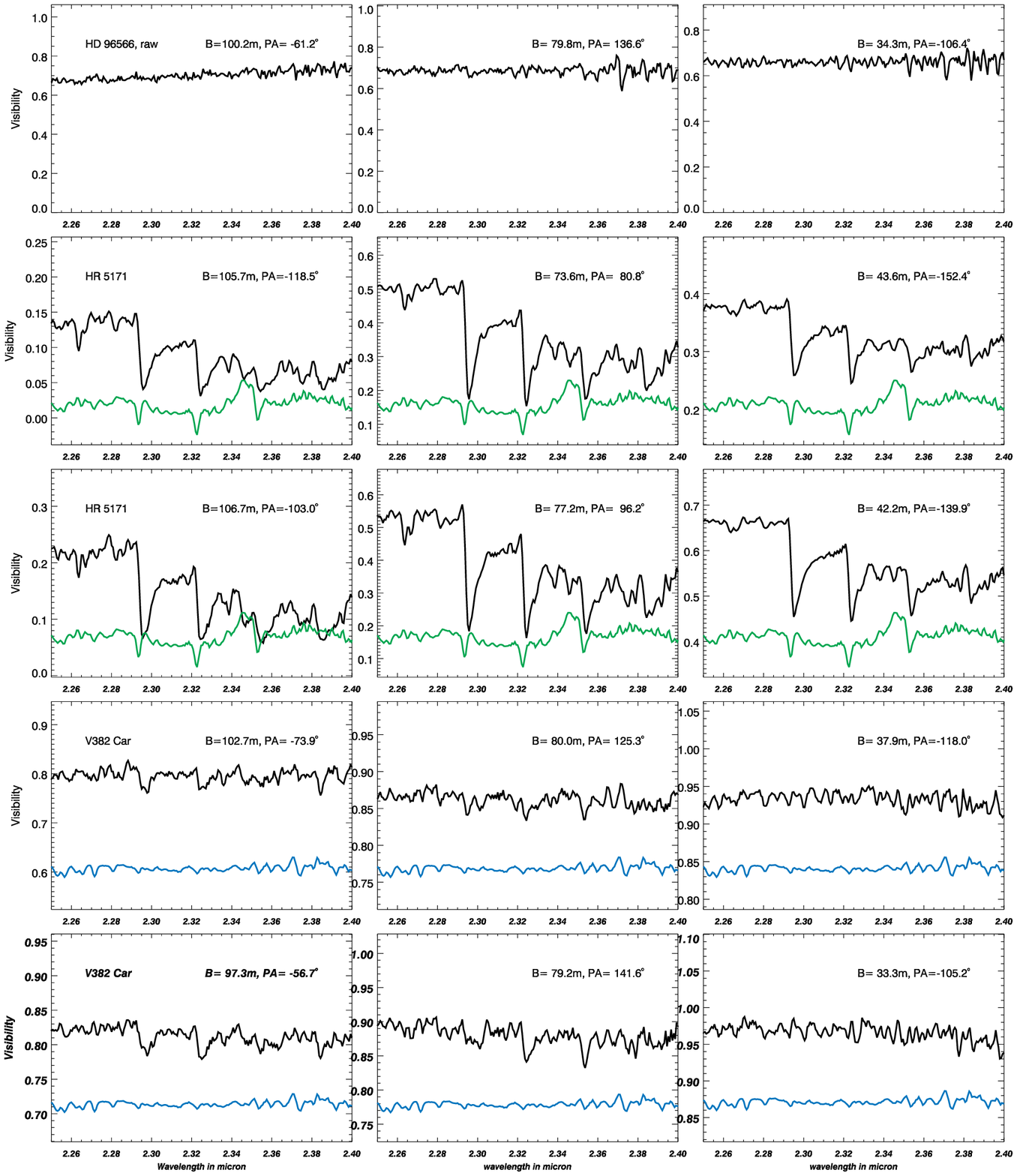}
 \caption[]{Medium-resolution (R=1500) AMBER/VLTI data centered in the CO molecular band. Some uncalibrated visibilities from a calibrator are shown in the upper row as a visual noise estimate. The dataset consists of three dispersed differential visibilities scaled to the level of the absolute visibility obtained using a calibrator. The spectrum (in arbitrary units) is shown in color at the bottom of each panel. The decrease in the dispersed visibilities betray some extended emission.    \label{fig:MRspec_CO_vis}}
\end{figure*}

\begin{figure*}
 \centering
\includegraphics[width=17.cm]{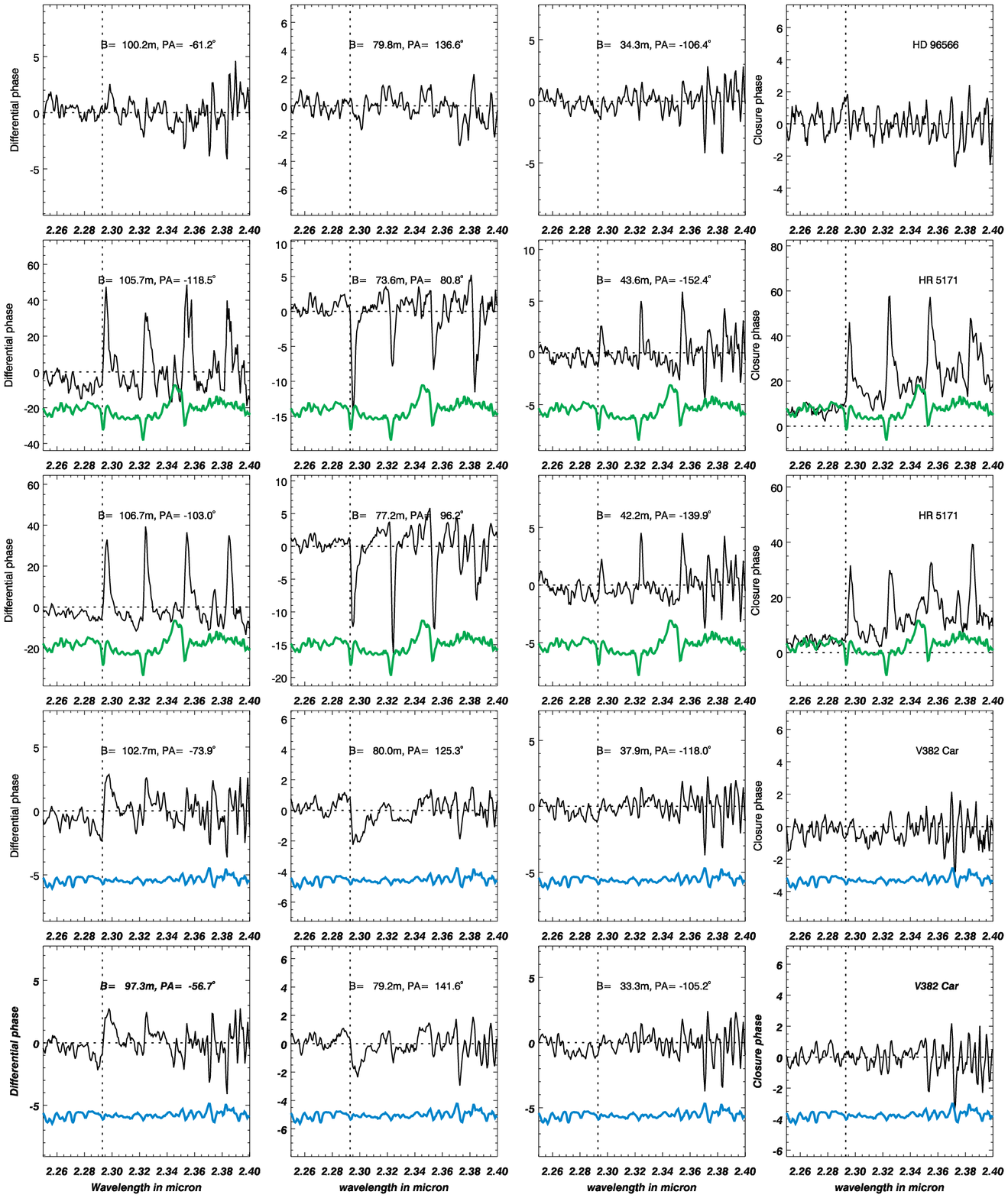}
 \caption[]{Same dataset and spectral region as that in the previous figure. The different panels show three differential phases and one closure phase per observation.   \label{fig:MRspec_CO_phas}}
\end{figure*}

\begin{figure*}
 \centering
\includegraphics[width=17.cm]{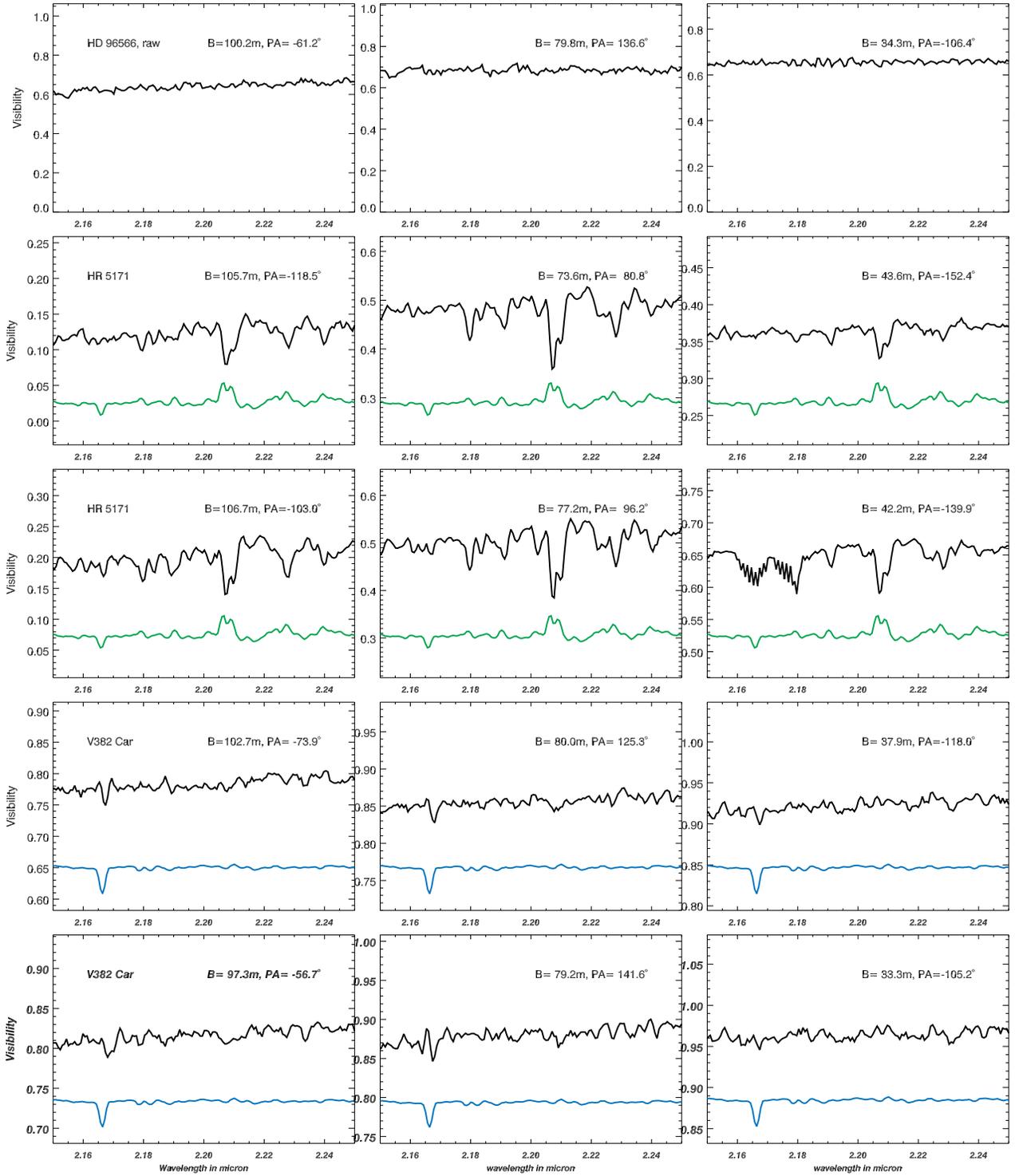}
 \caption[]{Medium-resolution (R=1500) AMBER/VLTI data centered on the Br$\gamma$ line. Some uncalibrated visibilities from a calibrator are shown in the upper row as a visual noise estimate. The dataset on the science consists of 5 consecutive observations providing for each, three dispersed visibilities (this figure), three differential phases and one closure phase (next figure). The decrease of the dispersed visibilities betray some extended emission, particularly in the NaI sodium line of \object{HR\,5171 A} whereas no signal is observed in the Br$\gamma$ line.  No significant interferometric signal is observed in the data from \object{V382\,Car}.   \label{fig:MRspec_Br_vis}}
\end{figure*}

\begin{figure*}
 \centering
\includegraphics[width=17.cm]{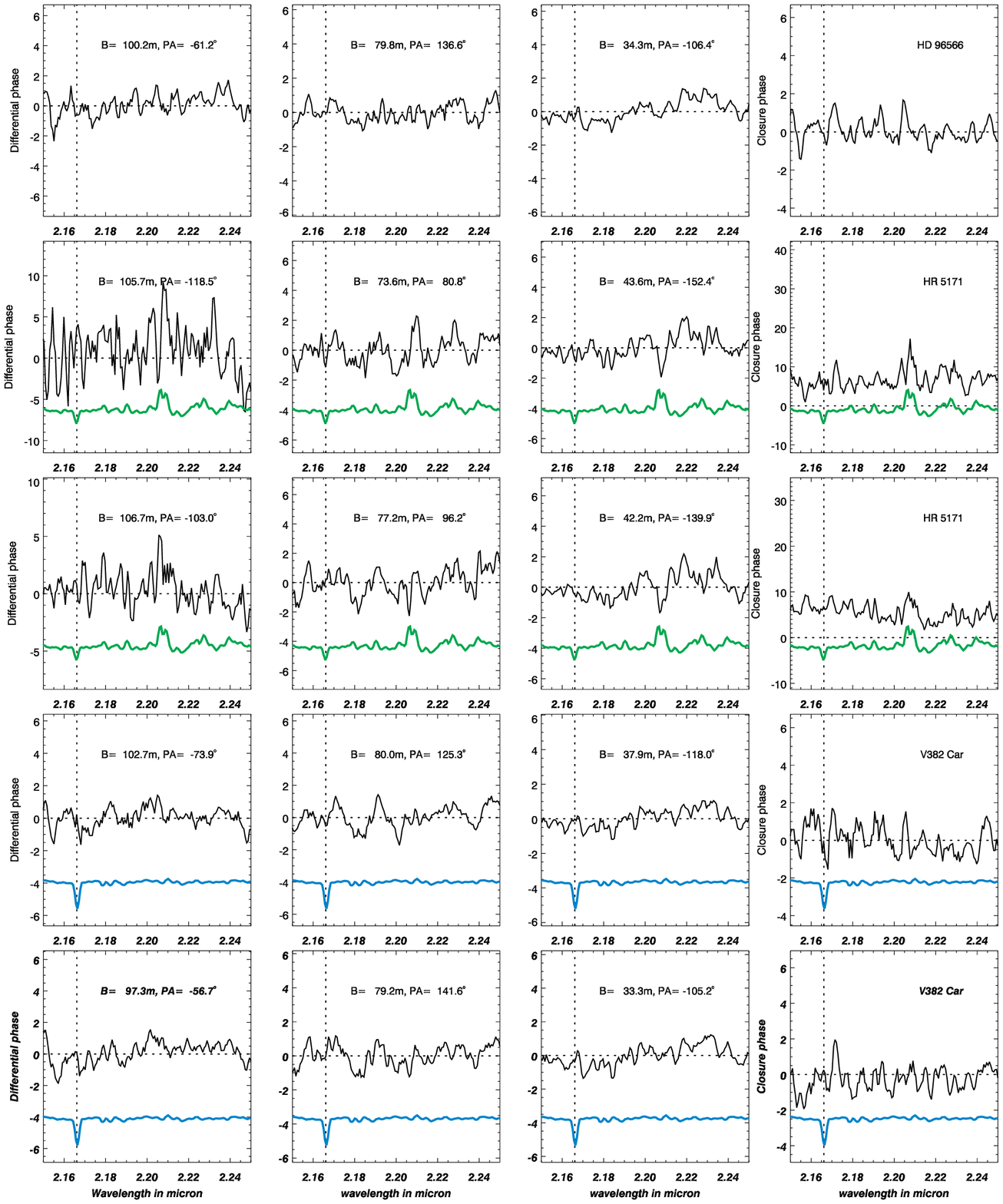}
 \caption[]{Same dataset and spectral region as that from the previous figure. The different panels show three differential phases and one closure phase per observation.    \label{fig:MRspec_Br_phas}}
\end{figure*}

The medium-resolution data from AMBER are very rich and deserve to be
shown in detail. The atmospheric conditions during the observing run
were excellent and the standard deviations of the dispersed
differential and closure phases are better than 2\deg. We show in
Fig.\,\ref{fig:MRspec_CO_vis} and in Fig.\,\ref{fig:MRspec_Br_vis} the
differential visibilities from five individual observations in the CO
and Br$\gamma$ region, respectively. The dataset consists of three
dispersed raw visibilities from a calibrator, and two sets of three
dispersed visibilities from \object{V382\,Car} and
\object{HR\,5171\,A} (out of five recorded each). The differential and
closure phases are shown in Fig.\,\ref{fig:MRspec_CO_phas} and
Fig.\,\ref{fig:MRspec_Br_phas} in the CO and Br$\gamma$ region,
respectively.

The comparison of the visibilities from these two sources shows the
strong impact of the extended environment of \object{HR\,5171\,A} on
the observations. We note also that a signal is observed in the
visibilities from \object{V382\,Car} through the CO lines, but only at
the longest baselines. This signal indicates a thin molecular
environment at a close distance from the photosphere. This must be some
CO emission filling an already existing CO absorption since there is
no significant signature of this CO in the spectrum. The strong phase
signal of the CO lines shows the spatial complexity of
\object{HR\,5171\,A}. We recall that a UD surrounded by a
centered GD should yield a zero phase signal. One can also see
that a significant closure phase signal is observed only in data from
\object{HR\,5171\,A}. Furthermore, the signal was observed to strongly
increase from one observation to the next during the night, following
the Earth rotation of the projected baselines, clear evidence of
binarity, but also an indication that the envelope of molecular gas also has a complex shape, not accounted for our geometrical model. No signal is observed in the Br$\gamma$ region (visibilities
and phases). A weak phase signal close to the noise limit is observed
in the NaI2.2$\mu$m sodium line of \object{HR\,5171\,A}.

\section{Archival X-ray data}
\label{Sec:X}

\object{HR\,5171\,A} and \object{HR\,5171\,B} were observed serendipitously by XMM-Newton during 40ks in Aug. 2001 (Rev number = 0315, thick filter used for the EPIC cameras). This archival dataset (ObsID=0087940201) was downloaded and processed using SAS v12.0.0 and calibration files available on July 1, 2012, following the recommendations of the {\tt XMM} team\footnote{SAS threads, see \\ http://xmm.esac.esa.int/sas/current/documentation/threads/ }. A background flare affects the beginning of the observation, which was discarded. HR\,5171 appears as a single faint source near the top edge of the field-of-view of the MOS2 and pn cameras. It is also near a gap in the pn dataset. The source detection was performed using the task {\it edetect\_chain} on the three EPIC datasets and in two energy bands (soft=S=0.3--2.0\,keV, hard=H=2.0--10.0\,keV energy band). The best-fit position for this source is 13:47:10.138,-62:35:16.11, i.e. at 8.6" to the NW of \object{HR\,5171\,A}, a position compatible with that of \object{HR\,5171\,B}. The X-ray detection is also compatible with current knowledge on the X-ray emission of massive stars : O- and early-B stars, such as \object{HR\,5171\,B}, are moderate X-ray emitters. We infer the equivalent on-axis count rates are 0.008 ct/s for MOS2 and 0.02 ct/s for pn, which correspond\footnote{conversion performed with HEASARC PIMMs, see http://heasarc.gsfc.nasa.gov/Tools/w3pimms.html} to fluxes of 1.5-2.55$\times 10^{-13}$\,erg.cm$^{-2}$\,s$^{-1}$ considering a plasma temperature of 0.6 keV. These values yield $\log(L_X/L_{BOL}) \sim -6.5$, a value typical of massive OB stars. In contrast, the evolved objects with strong winds, such as the LBVs or WCs, are generally not detected \citep{2009A&A...506.1055N,2012A&A...538A..47N,2003A&A...402..755O}. It is thus unsurprising that  \object{HR\,5171\,A} is not detected. There is also no evidence of a companion or any evidence of energetic radiations from the interaction between the two components in this spectral range. In August 2001 the orbital phase was about 0.66, a potentially favorable configuration for observing any emission from the companion.

\section{Infrared photometry}
\label{sec:midIR}

\object{HR\,5171\,A} shows dust excess \citep{1971ApJ...167L..35H, 1975ApJ...196..761A, 1986ApJ...307..711O, 1985Obs...105..229S}. Since the early 70s, \object{HR\,5171\,A} has regularly been  observed by space-borne infrared instruments, and we report these measurements in Table\,\ref{tab:sed}. The aperture of these instruments is larger than several arcseconds. Since \object{HR\,5171\,A} cleaned up a large cavity, one can reasonably be confident that no other infrared source contributes significantly to the infrared flux. The flux from the B0Ia supergiant \object{HR\,5171\,B} is also negligible at these wavelengths. Some significant variability of the source is observed by comparing the MSX and WISE measurements obtained $\sim15$yr apart in filters with relatively close characteristics. The recent WISE fluxes are systematically lower at wavelengths shorter than 12\micron. Nonetheless, the large variations observed between instruments with filters covering the N band can be explained by the variations of the transmission that include different contributions of the very strong silicate features at 10 $\mu$m.

\begin{figure*}
 \centering
\includegraphics[width=9.cm]{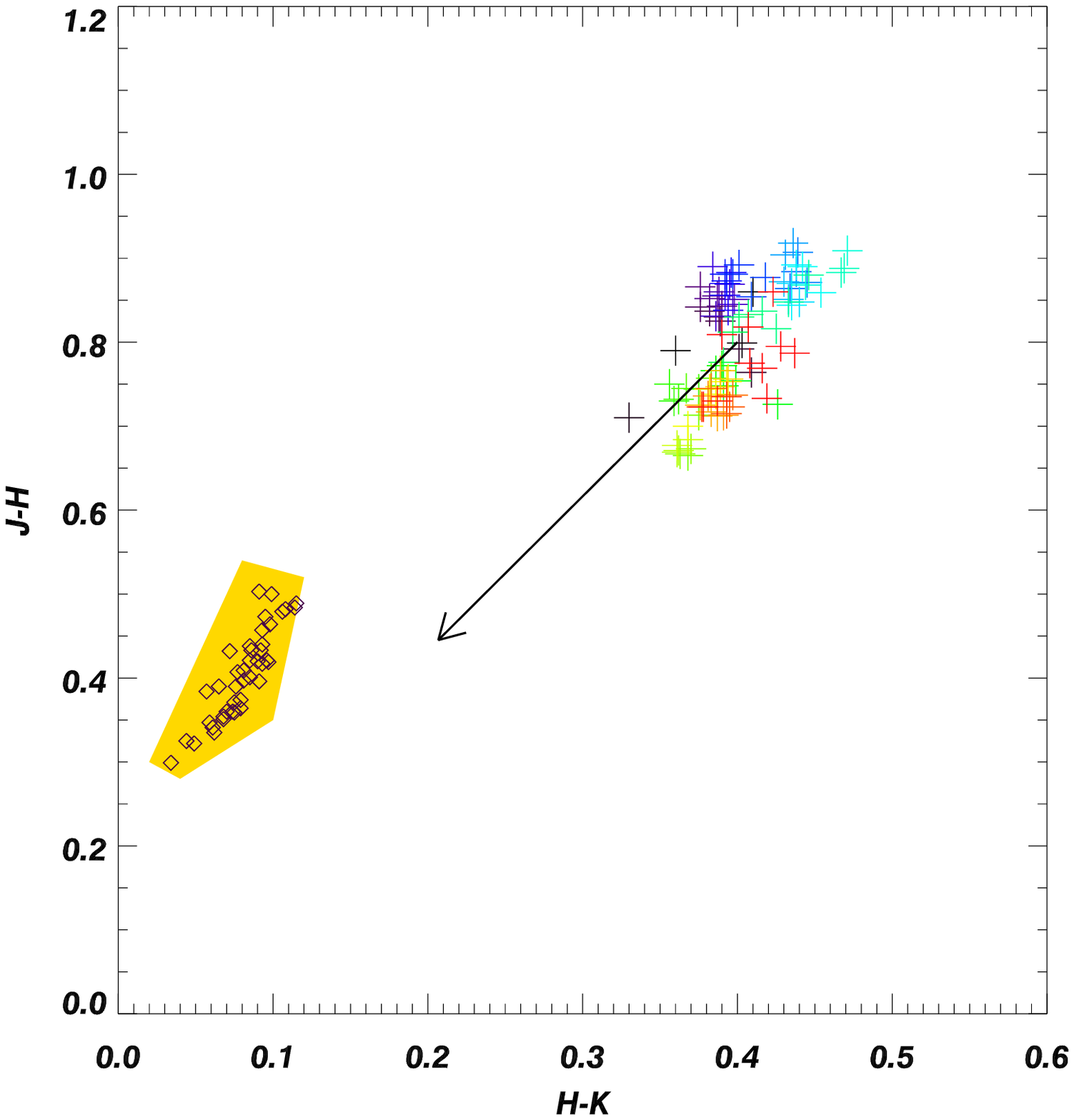}
\includegraphics[width=9.cm]{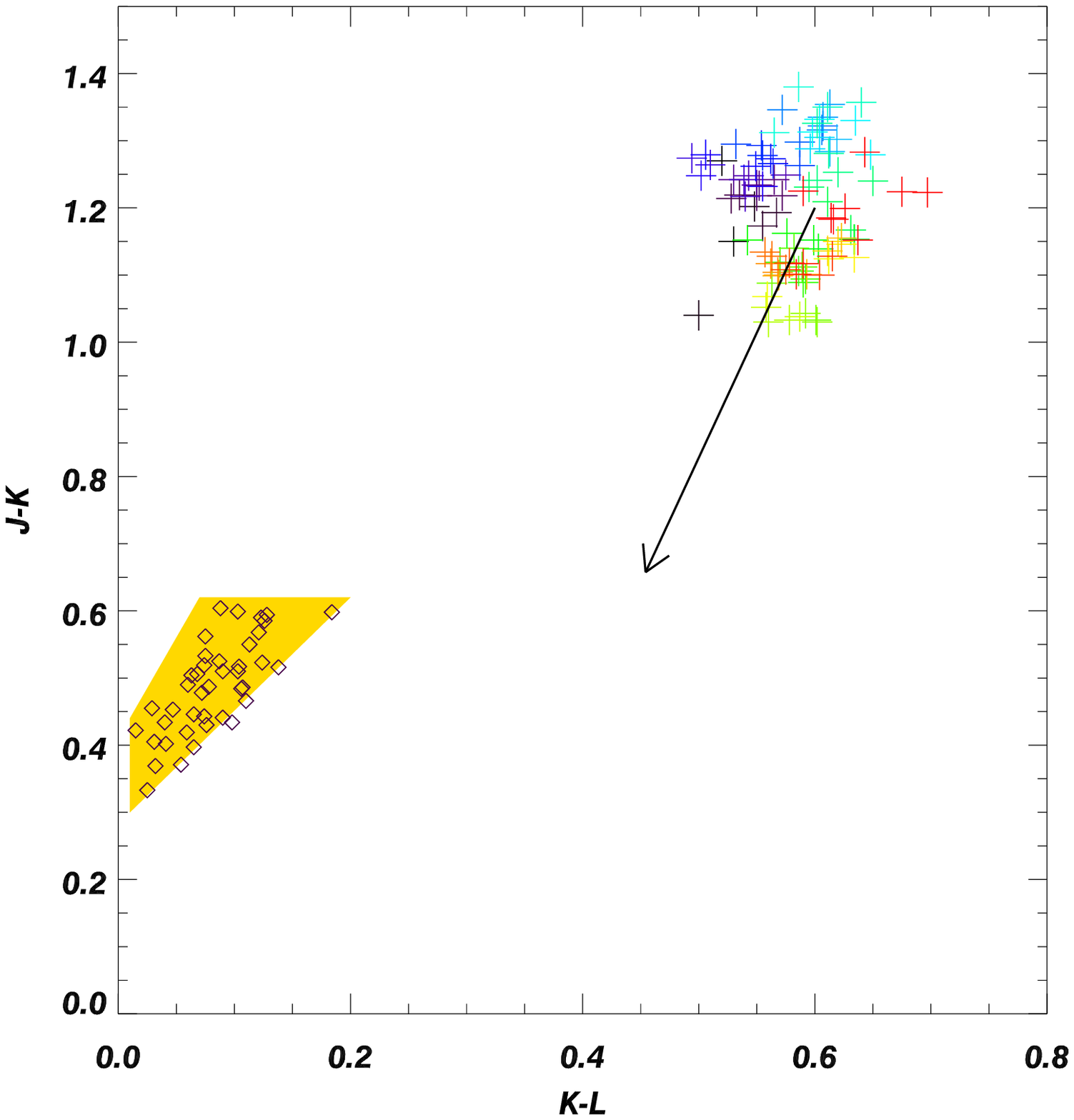}
 \caption[]{Near-infrared two colour diagrams. The full dataset is shown with the same color coding as in Fig.\ref{fig:nearColors} and the location of Cepheids \citep{1992ASPC...30..119L}, after correction for interstellar reddening, [assuming the reddening law from \citet{1993MNRAS.263..921L} and A$_L$=0.15\,E(B-V)] is shown for comparison. The arrow shows the effect of correcting HR5171A for interstellar reddening of E(B-V)=1.13 mag \citep{1992A&A...257..177V}, following the same reddening law. We note that there may well be additional reddening due to circumstellar extinction of uncertain characteristics, evidenced by the spread (0.1-0.3 mag) of the temporal measurements.  Nevertheless, the  corrected colors are similar to those of the Cepheids, with an excess  of emission that increases at longer wavelengths. \label{fig:jh-hk}}
\end{figure*}

The $JHKL$ colors of \object{HR\,5171\,A} are compared to those of YSGs, i.e. Cepheid variables in Figs.\,\ref{fig:jh-hk}. They are very similar in $JHK$, but \object{HR\,5171\,A} has distinctly redder colors in
$K-L$, probably from the influence of circumstellar dust.

\begin{table} [h]
  \begin{caption}
    {Photometry of the total source. 
     \label{tab:sed}
    } 
  \end{caption}
  \centering
  \begin{tabular}{lcccc}
    \hline
   Instrument & $\lambda_0$ & Filter & Flux  & Date \\
   & [micron] &  & [Jy] &  \\
        \hline
   RAFLGL & 12.2 & - & 692 & 1971\\
   RAFLGL & 20 & - & 789 & 1971\\
    \hline
   IRAS & - & 12 & 605 & 1983\\
   IRAS & -& 25 & 531 & 1983\\
   
     \hline
    MSX$^{\mathrm{1}}$ & 4.29 & B1 & 179$\pm$15 & 1996/1997 \\
    MSX & 4.25 & B2 & 230$\pm$20 & 1996/1997\\
    MSX & 8.9 & A & 176$\pm$7 & 1996/1997\\
    MSX & 12.1 & C & 679$\pm$34 &1996/1997\\
    MSX & 14.6 & D & 516$\pm$31 &1996/1997\\
    MSX & 21.3 & E & 703$\pm$49 &1996/1997\\
       \hline
	AKARI$^{\mathrm{2}}$ & 18 & S18 &1121$\pm$182 & 2006/2007\\
	AKARI$^{\mathrm{3}}$ & 65 & S65&95$\pm$13 & 2006/2007\\
	AKARI & 90 & S90&37$\pm$5 & 2006/2007\\
	AKARI & 140 & S140 &11$\pm$6 & 2006/2007\\
     \hline
		WISE$^{\mathrm{4}}$ &   3.4 & W1 & 95$^{\mathrm{5}}$ & 2010 \\
 		WISE &   4.6 & W2 & 67$\pm$6 & 2010 \\
		WISE &   12 & W3 & 291$\pm$39 & 2010 \\
		WISE &   22 & W4 & 806$\pm$7 & 2010 \\
      \hline

 \end{tabular}
 	\begin{list}{}{}
 	\item[$^{\mathrm{1}}$] \citet{2003yCat.5114....0E}
  	\item[$^{\mathrm{2}}$] \citet{2010A&A...514A...1I}
 	\item[$^{\mathrm{3}}$] \citet{2010yCat.2298....0Y}
	\item[$^{\mathrm{4}}$] \citet{2012yCat.2311....0C}
 	\item[$^{\mathrm{5}}$] The exposure is saturated.
	\end{list}

 \end{table}

%Near-IR color-color diagrams, $JH$ versus $HK$ on the left and $JK$ versus $KL$ on the right. The location of Cepheids %is indicated as dashed bow together with some measurements from \citet{1992ASPC...30..119L} as diamonds. The full %dataset is shown with the same color coding as in Fig.\ref{fig:nearColors}. The arrows indicate a lower limit for the %interstellar reddening based on the observations of the visual companion \object{HR\,5171\,A} with an excess %$E(B-V)=0.7-0.8$ . An additional circumstellar component ranging between 0.1 and 0.3 magnitude must be taken into %account, whose variable part explains well the particular shape of the cloud formed by the  location of %\object{HR\,5171\,A} versus time in the color-color diagrams.  

\end{document}